\documentclass[sigconf]{acmart}

\usepackage{graphicx}
\usepackage{enumitem}
\usepackage{balance}
\usepackage{multirow}
\usepackage{booktabs}
\usepackage{xspace} 
\usepackage{amsmath,amsthm,amsfonts}

\usepackage{amssymb}

\newcommand{\meta}[1]{{\color{brown}\{\{}{\color{blue}#1}{\color{brown}\}\}}}

\AtBeginDocument{%
  \providecommand\BibTeX{{%
    \normalfont B\kern-0.5em{\scshape i\kern-0.25em b}\kern-0.8em\TeX}}}

\setcopyright{acmcopyright}
\copyrightyear{2018}
\acmYear{2018}
\acmDOI{XXXXXXX.XXXXXXX}

\acmConference[Conference acronym 'XX]{Make sure to enter the correct
  conference title from your rights confirmation emai}{June 03--05,
  2018}{Woodstock, NY}
\acmPrice{15.00}
\acmISBN{978-1-4503-XXXX-X/18/06}

\begin{document}


\title{Knowledge Plugins: Enhancing Large Language Models for Domain-Specific Recommendations}

\author{Jing Yao$^1$, Wei Xu$^{2}$, Jianxun Lian$^1$, Xiting Wang$^1$, Xiaoyuan Yi$^1$ and Xing Xie$^1$}
\authornote{Work done during Wei Xu's internship at Microsoft Research Asia.}
\affiliation{%
  $^1$Microsoft Research Asia, Beijing, China \\
  $^2$Gaoling School of Artificial Intelligence, Renmin University of China \\
  \country{}
}
\email{{jingyao, jianxun.lian, xiting.wang, xing.xie}@microsoft.com}
\email{xu\_wei@ruc.edu.cn}
\settopmatter{printacmref=true}

\begin{abstract}
The significant progress of large language models (LLMs) provides a promising opportunity to build human-like systems for various practical applications.
However, when applied to specific task domains, an LLM pre-trained on a general-purpose corpus may exhibit a deficit or inadequacy in two types of domain-specific knowledge. One is a comprehensive set of domain data that is typically large-scale and continuously evolving. The other is specific working patterns of this domain reflected in the data. The absence or inadequacy of such knowledge impacts the performance of the LLM.
In this paper, we propose a general paradigm that augments LLMs with \underline{DO}main-specific \underline{K}nowledg\underline{E} to enhance their performance on practical applications, namely \underline{DOKE}. This paradigm relies on a domain knowledge extractor, working in three steps: 1) preparing effective knowledge for the task; 2) selecting the knowledge for each specific sample; and 3) expressing the knowledge in an LLM-understandable way. Then, the extracted knowledge is incorporated through prompts, without any computational cost of model fine-tuning.
We instantiate the general paradigm on a widespread application, i.e. recommender systems, where critical item attributes and collaborative filtering signals are incorporated.
Experimental results demonstrate that DOKE can substantially improve the performance of LLMs in specific domains.
\end{abstract}



\maketitle

\section{Introduction}
Large language models (LLMs) have made significant progress recently~\cite{brown2020language,ouyang2022instructgpt,chowdhery2022palm,zhang2022opt,zeng2022glm,touvron2023llama}. With more than billions of parameters, LLMs emerge with remarkable capabilities, such as in-context learning, complex reasoning, and instruction following~\cite{wei2022emergent}. Unlike conventional deep models, which are tailored for specific tasks and datasets, LLMs endowed with rich internal knowledge exhibit superior generalization to novel tasks. This can effectively address many challenges encountered by conventional models, such as cold-start scenarios that lack sufficient training data.
Therefore, harnessing LLMs to build versatile and human-like systems for various applications emerges as a new research direction, which has great potential to revolutionize the corresponding application, such as recommender systems stated in~\cite{wu2023llm_rec_survey,fan2023llm_rec_survey}.

\begin{figure}
    \centering
    \includegraphics[width=0.99\linewidth]{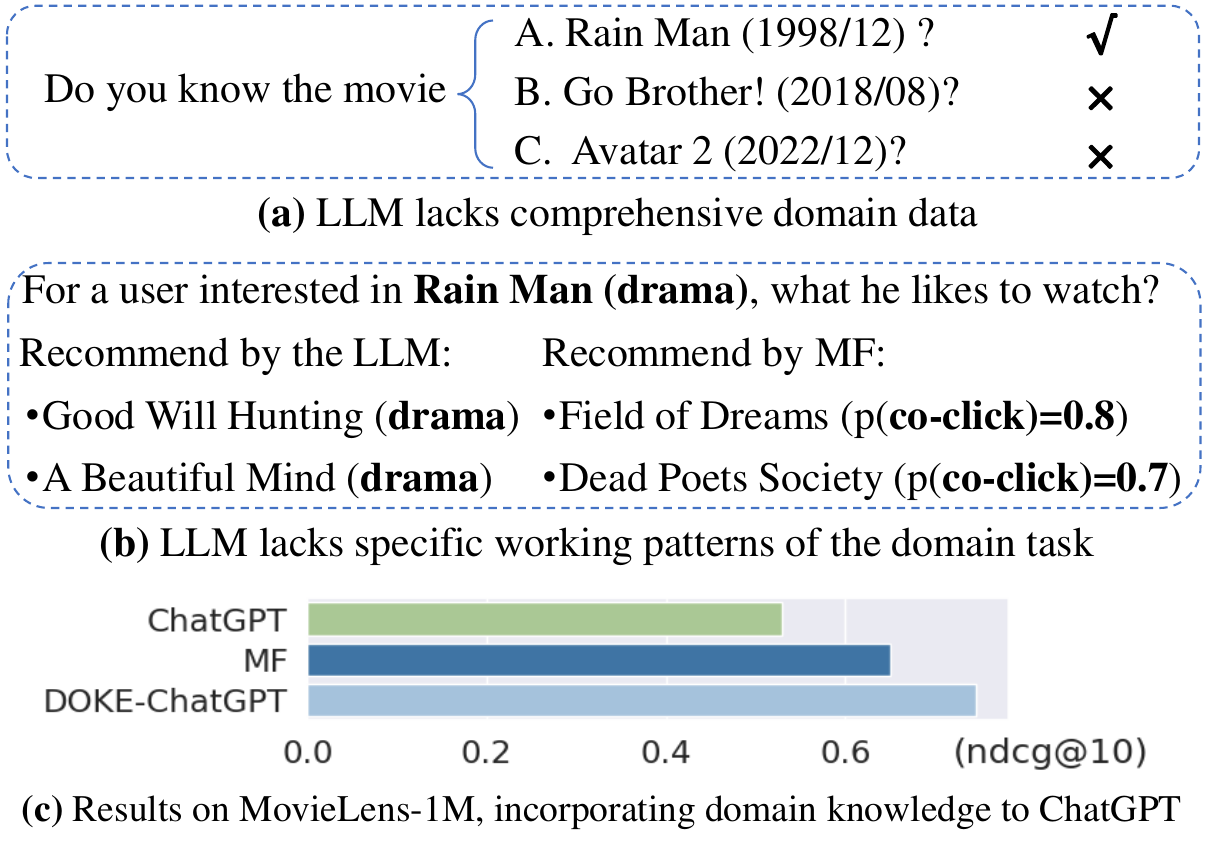}
    \caption{An illustration of domain knowledge lacked in the LLM for movie recommendation.}
    \label{fig:example}
    \vspace{-4.2mm}
\end{figure}

Since an LLM is pre-trained on a general text corpus, we argue that it may exhibit a deficit or inadequacy in two types of domain-specific knowledge when directly applied to a specific domain. This prevents it from achieving good performance while possessing other unique advantages. 
On the one hand, it is difficult for an LLM to memorize full domain data within its parameters, because the dataset is typically enumerou and continuously evolving. On the other hand, an LLM may only have rough and general experience about how to accomplish a task, but the knowledge about professional working patterns of the specific domain is still contained in the collected domain records. These records may be less accessible during the pre-training stage.
Figure~\ref{fig:example} presents an example of a movie recommendation task to illustrate the two issues. First, the LLM clearly recognizes some well-known movies (such as \emph{Rain Man}) in this domain but not all, especially those less popular and fresh items (such as \emph{Avatar 2}). Furthermore, we observe that the LLM might make recommendations using a general rule, i.e. relevance between some textual features or descriptions. For example, it recommends \emph{Good will Hunting} for the user interested in \emph{Rain Man} due to their sharing genre of `drama'. But sometimes this is not consistent with the observed user behaviors in a specific domain. More valuable signals for recommendation beyond semantic similarity are reflected in the domain-specific interaction data~\cite{kang2018sasrec,he2017neuralcf}.
If necessary knowledge for making recommendations in this specific domain is incorporated into the LLM, such as the co-click probability of items, better performance can be obtained.

To augment LLMs with factual and up-to-date world knowledge to alleviate hallucination, many retrieval-augmented methods~\cite{liu2021retrievalllm,yu2023retrievalcorpus} are proposed, but they do not consider domain-specific knowledge that varies across tasks and domains. 
In addition, fine-tuning model parameters on the target dataset is a widely used approach to adapt previous language models such as BERT~\cite{devlin2018bert} for a specific domain~\cite{geng2022p5,cui2022m6_rec,he2020parade_finetune}. However, this solution is expensive for LLMs with massive parameters due to the requirement of huge computational and storage costs. As the model size increases, it would become more infeasible. Moreover, fine-tuning \#with domain-specific data\# is prone to overfitting and may sacrifice general intelligence to deal with other challenges~\cite{luo2023catastrophic_forget}.

In this paper, we present a general paradigm, namely \textbf{DOKE}, that augments LLMs with \underline{DO}main-specific \underline{K}nowledg\underline{E} to enhance their performance in practical domains.
This paradigm aims to involve any knowledge that is contained in the domain data and beneficial for the task, and then it incorporates this knowledge through prompts, without any modification of LLM parameters. It mainly relies on an external domain knowledge extractor that works in three steps. (1) \emph{Knowledge Preparation}: it first fully understands the domain data and extracts domain knowledge necessary for this task. (2) \emph{Knowledge Customization}: since we should incorporate the knowledge through prompts with a limited length, it only selects domain knowledge relevant to the current sample. (3) \emph{Knowledge Expression}: it finally converts the knowledge into natural language that can be understood and accepted by all LLMs. In addition, this external knowledge extractor can be any lightweight model specified for the domain. Thus, our proposed augmentation paradigm is general across various domains to provide professional knowledge and is much less costly than those fine-tuning approaches.

To verify our proposed paradigm, we present an instance on a widespread AI application, i.e. recommender system, where LLMs lack domain-specific knowledge to acquire great accuracy~\cite{hou2023large_rec_eval,liu2023chatgpt_rec_eval}, as shown in Figure~\ref{fig:example}. Since the pre-trained LLM suffers from memorizing the large-scale and evolving item set, we formalize the task as ranking a small set of candidate items retrieved by an off-the-shelf method. Then, we devise a domain knowledge extractor to obtain information that is beneficial for the recommendation task, including item attributes and collaborative filtering (CF) signals among users. We customize the domain knowledge for each sample based on the current user's preferences and candidate items. Moreover, we consider two expressive formats to translate the provided information to the LLM, i.e. natural language templates and reasoning paths on a knowledge graph (KG) that are more explainable~\cite{xian2019kgexplainable}.
In the paradigm of building human-like recommender systems with LLMs, experimental results show that DOKE can significantly improve the performance of LLMs by incorporating high-quality domain-specific knowledge.

The main contributions are summarized as follows:

$\bullet$ We explore that LLMs lack domain-specific knowledge to achieve great performance in practical domains. To alleviate this issue, we propose a general augmentation paradigm with a domain knowledge extractor, which is applicable across various domains more cost-effective than fine-tuning approaches.


$\bullet$ We showcase an example of the paradigm on recommender systems. The domain knowledge extractor provides item attributes and CF information, which is customized for each specific sample and expressed in natural language or reasoning paths on KG.

$\bullet$ We conduct experiments on recommendation benchmarks from different domains. Results show that our paradigm can achieve the best performance among zero-shot LLMs and is comparable to full trained models on specific domains.


\begin{figure*}[ht]
    \centering
    \includegraphics[width=0.95\linewidth]{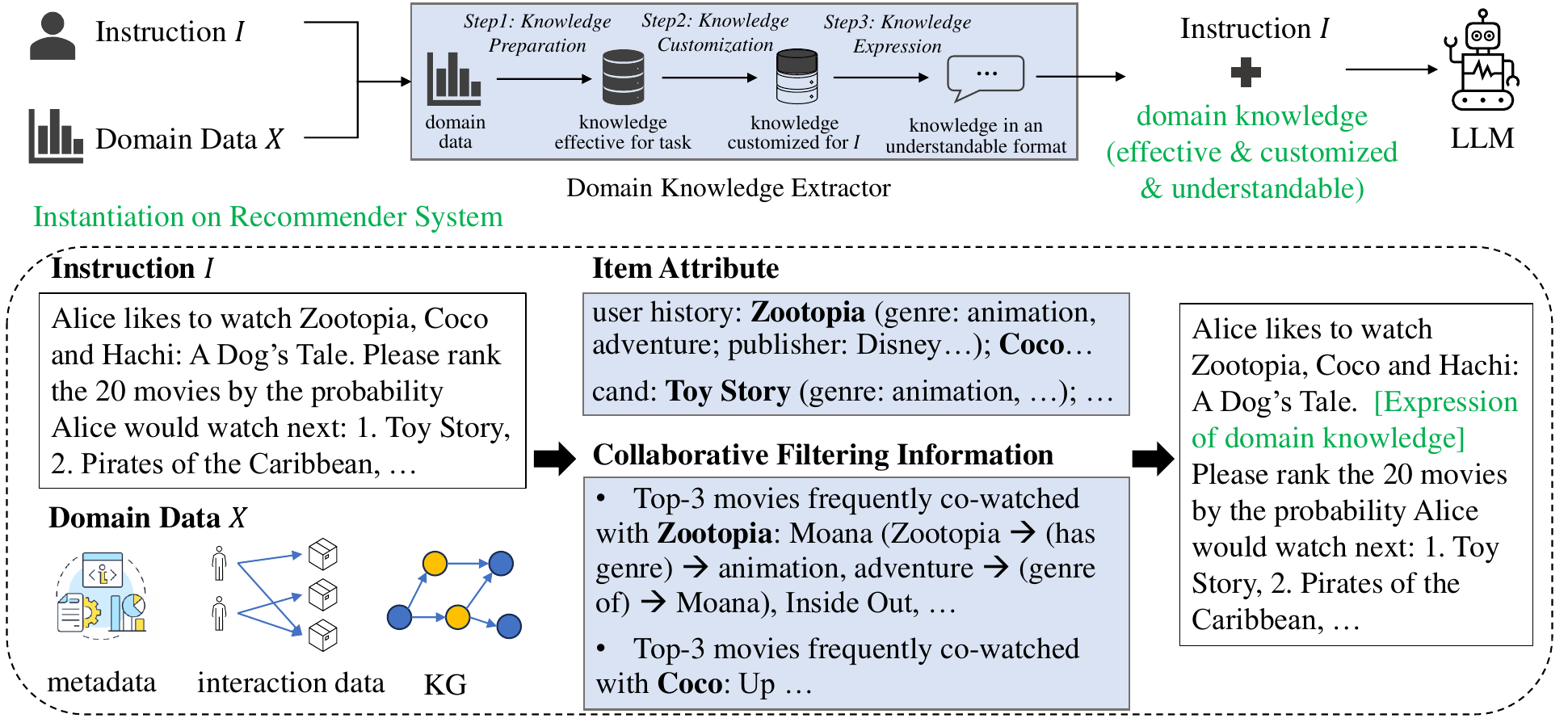}
    \caption{Illustration of our DOKE paradigm for LLMs with a domain knowledge extractor that works in three steps: 1) preparing effective knowledge for the task; 2) customizing knowledge for the current sample; and 3) expressing knowledge in natural language. In the instantiation on recommender systems, item attributes, and CF information are extracted. They are customized based on the user and candidate items, expressed by natural language and knowledge reasoning paths.}
    \label{fig:framework}
\end{figure*}

\section{Related Work}
\subsection{LLMs for Practical Applications}
The impressive capabilities of LLMs inspire researchers to extend their scope to more practical applications, such as recommender systems. In the early stage, some studies~\cite{xi2023llm_enhance_kg,lyu2023llm_enhance_kg,du2023llm_enhance_query,li2023llm_enhance_query,wu2021llm_enhance_encode,hou2022llm_enhance_encode,xiao2022llm_enhance_encode} incorporates LLMs to enhance their original models, which plays a role of text encoder, feature extractor or external knowledge base. As LLMs become more intelligent, another kind of approach sets the LLM as a task manager and allows it to use tools for completion~\cite{schick2023toolformer,gao2023chatrec,liang2023taskmatrix}, such visual-chatgpt~\cite{wu2023visual_chatgpt,shen2023hugginggpt}. For the two categories of methods, the final results are still determined by the conventional models, preventing LLMs' potential.
Furthermore, the new paradigm of directly harnessing powerful LLMs for various applications receives growing interest. First, some works construct task-specific prompts to apply LLMs~\cite{hou2023large_rec_eval,liu2023chatgpt_rec_eval}, where different prompt strategies such as in-context learning (ICL) and chain-of-thought (CoT) are mainly explored~\cite{wang2023prompt_rec,kang2023prompt_example_finetune,hou2023prompt_icl_example}. This naive approach confirms that LLMs have promising zero-shot capabilities, but still lag far behind neural models fully trained on the domain data. To bridge this gap and introduce domain-specific knowledge, a kind of approach finetunes a general pre-trained model with the target data~\cite{bao2023tallrec,zhang2023recommendation}, such as fine-tuned recommender systems~\cite{zhang2021language_rec,geng2022p5,cui2022m6_rec}. However, this solution is resource-intensive for LLMs with large-scale parameters, because much computation and storage cost is required. And fine-tuning on specific domain data is prone to overfitting and may sacrifice general intelligence.

In this paper, we focus on the new paradigm of developing LLMs as human-like systems for practical applications. Different from fine-tuning approaches, we design a paradigm to obtain effective information and incorporate it through prompts. Thus, we can fully leverage the power of LLMs and eliminate the fine-tuning cost.

\subsection{Retrieval-Augmented LLMs}
There have been a lot of works exploring how to augment LLMs to extend the scope of inner knowledge and deal with the issue of hallucination. Inspired by the powerful in-context learning ability of GPT-3, Liu et. al.\cite{liu2021incontext} retrieve semantically similar examples to enhance the few-shot performance. Many retrieval-augmented approaches are proposed for QA and complex reasoning tasks. Given a question, factual and up-to-date documents or knowledge are retrieved from a collected corpus~\cite{yu2023retrievalcorpus,izacard2022retrievalcorpus}, a search engine~\cite{lazaridou2022retrievalinternet}, or a pre-trained language model~\cite{liu2021retrievalllm}. Then, LLMs generate the final answer based on the original question and the incorporated facts.

These works focus on appending general and real-time facts for open questions to eliminate hallucinations, while our paper is more concerned with the lack of domain-specific knowledge that varies across tasks and domains.

\subsection{Classical Recommendation Methods}
Recommender systems have made significant progress in the past decades~\cite{das2017rec_survey}. 
In the early stage, user preference modeling and recommendations mainly rely on item content~\cite{kobsa1994contentrec}. Later, collaborative filtering (CF) algorithms~\cite{resnick1994usercf,sarwar2001itemcf} predict user interactions based on associative behaviors, where the similarity matching between users and items is critical. Furthermore, approaches such as MF~\cite{koren2009matrixfactor} and NCF~\cite{he2017neuralcf} learn latent vectors for users and items. Sequential recommendation learns better user preferences by focusing more on the closer interaction history~\cite{kang2018sasrec}. Knowledge graph-based recommendation ~\cite{wang2019kgexplainable,xian2019kgexplainable,zhao2020kgexplainable,liu2021kgrec} incorporates external knowledge to alleviate data sparsity and improve recommendation explainability.

These recommendation methods are typically designed for specific tasks and fit to a given data distribution, while LLMs-based recommenders could yield human-like capabilities and better generalization for novel tasks and datasets. Whereas, they can capture professional knowledge from the data to perform well in the specific domain, such as item content and relevance between users and items. We attempt to introduce this knowledge to enhance the domain-specific performance of LLMs.

\section{DOKE: The General Paradigm}
The entire workflow of our proposed paradigm for \underline{DO}main-specific \underline{K}nowledg\underline{E} enhancement, namely DOKE, is illustrated in Figure~\ref{fig:framework}. Suppose there is a task from a specific domain $D$ which has a corresponding set of training data $\mathcal{X}$, a direct approach of applying LLMs to solve this task is first formulating it as an instruction $I$ and prompting LLMs to generate the result, without any proactive adjustments. Under our DOKE paradigm, we devise a domain knowledge extractor $F^D$ to obtain high-quality knowledge from the domain data for the current instruction $I$, denoted as $K^D_I$. Then, the extracted knowledge is combined with the original instruction as the final prompt for the result generation. Here, the core component of this paradigm is the domain knowledge extractor, which works in the following three steps:

\emph{Step1: Knowledge Preparation.} To obtain high-quality information for enhancement, the extractor is expected to fully understand the domain data, identify what information is necessary for completing the task and successfully extract useful domain-specific knowledge.

\emph{Step2: Knowledge Customization.} Since the available domain knowledge extracted from the whole data can be rich and broad, it is impossible to directly append all in each prompt due to the limitation on length. Besides, selecting the knowledge most relevant to the current sample but ignoring other irrelevant could be more effective and efficient. Thus, the extractor tailors the knowledge to be incorporated for each specific sample.

\emph{Step3: Knowledge Expression.} The domain knowledge beneficial for the task can be in various data formats, for example, the co-click rate is a float number as shown in Figure~\ref{fig:example}. Natural language is the unified way for LLMs to receive and process information. Furthermore, LLMs would benefit from in-context learning and chain-of-thought~\cite{wei2022chain}. To allow LLMs to better understand and benefit from the provided knowledge, the extracted knowledge is expressed in an easily understandable and rational format.

The domain knowledge extractor could make use of some task-specific models that are lightweight enough and well-trained in this domain. Their training cost is much less compared to that of fine-tuning LLMs, and the latency can be ignored compared to the inference time of LLMs. In this way, DOKE becomes applicable across various domains and is much less costly than fine-tuning approaches. In the next section, we showcase an instantiation of DOKE for recommender systems, illustrated in Figure~\ref{fig:framework}.

\section{DOKE for Recommender System}
In this paper, we focus on developing human-like LLMs for recommendation, and the problem is first defined as follows.
In a recommender system of a specific domain, there is a user set $\mathcal{U}$, an item set $\mathcal{I}$ and an interaction matrix $Y \in R^{|U|\times |I|}$. Besides, attribute information of both users and items, i.e. $\mathcal{A^{U}}$ and $\mathcal{A^{I}}$, is usually available, and sometimes a domain knowledge graph $\mathcal{G}$ is also maintained. All these constitute the whole domain data $\mathcal{X} = \{\mathcal{U}, \mathcal{I}, Y, \mathcal{A^{U}}, \mathcal{A^{I}}, \mathcal{G}\}$. 

When making recommendations for a user $u$, an existing recommender system first retrieves a small number of items from the whole set and then ranks these candidates according to their click probabilities. Since the item set is large-scale and continuously evolving, it is infeasible for the LLM to memorize and track the entire item set for retrieval. Therefore, we retrieve items with existing methods and then exploit the LLM to rank candidate items for the final recommendation result. This task is formalized as:
given the interaction history of a user $H^u = \{h^u_1, h^u_2,\ldots, h^u_{l}\}$, a set of candidate items $C=\{i_1, i_2, \ldots\, i_{|C|}\}$ is first retrieved using an off-the-shelf method. With the user's interaction history $H^u$ and the candidate item set $C$ as the prompt, the LLM generates a ranking list of these candidates.

Under our proposed DOKE paradigm, domain knowledge is extracted from the available data $\mathcal{X}$ through the three steps and appended to the above prompt. More details about domain knowledge extraction are introduced in the next.

\subsection{Knowledge Preparation}
Referring to classical recommendation models and the primary missing information of LLMs, we mainly consider two widely concerned types of domain knowledge to assist LLMs with the above-defined recommendation task.

$\bullet$ \textbf{Item Attribute}. Due to the entire item set is typically large-scale and continuously evolving, and some recommendation domains are relatively closed with less publicly available knowledge, such as the information of some products on Amazon, the LLM can suffer from clearly recognizing all items. Thus, the most basic domain knowledge we should provide is the attribute information of historical and candidate items, based on which the LLM can leverage its inner knowledge to understand the item better. For example, given a fresh product on Amazon with the title, brand, category, price, and other detailed attributes, the intellectual LLM can identify the characteristics of this item. Furthermore, item attributes have been proven to be informative in existing recommendation methods~\cite{wu2019attention}.


$\bullet$ \textbf{Collaborative Filtering Information}. In addition to basic attributes, collaborative filtering (CF) information is another widely used signal in recommendation methods. Since accurate CF signals on a domain can only be mined from the interaction records which are treated as private data for both the recommender system and users, this kind of knowledge could be lacking in the LLM. 
CF information identifies popular patterns and relevance between users and items. This is grounded in massive real-user interactions and truly reflects the preference patterns of users in this specific domain. By incorporating CF information with high confidence to the LLM, it can learn specific recommendation patterns from the interaction data, rather than just relying on general content associations inference. We consider providing CF information from two main perspectives.

(1) Item-2-Item: We think the correlation between two items can reflect the most basic recommendation patterns. For example, if many pairs of products co-bought by users belong to the same brand, we can learn that making recommendations based on the brand feature is effective in this domain. We follow the item-based CF method to calculate the correlation between a pair of items $(i_1, i_2)$ as:
\begin{equation}\label{eq:item_rel}
    rel(i_1, i_2) = \frac{\text{co-occur}(i_1, i_2)}{sqrt(N_{i_1} \times N{i_2})},
\end{equation}
where $N_{i_1}$ and $N_{i_2}$ are their click frequencies respectively, and $\text{co-occur}(i_1, i_2)$ denotes how many times they are co-clicked by the same user. Besides, some models such as MF learn the representation vector of items (e.g. $v_i$), and we can compute the inner product to gain more generalized relevance.
\begin{equation}\label{eq:item_vec_rel}
    rel(i_1, i_2) = v_{i_1} \cdot v_{i_2}.
\end{equation}
We provide the LLM with item pairs that have high relevance scores to indicate domain knowledge.

(2) User-2-Item: More general recommendations are based on the relevance between users and items, taking into account the whole user interaction history instead of a single item. This shows more comprehensive patterns between user preferences and items in this domain. Many methods learn user and item representations from the interaction data, i.e. $v_u, v_i$. We can employ any such recommendation model to compute the relevance between a user $u$ and an item $i$ by the inner product:
\begin{equation}\label{eq:user_item_rel}
    rel(u, i) = v_{u} \cdot v_{i}.
\end{equation}
In this paper, we adopt the SASRec model~\cite{kang2018sasrec}. Items with high relevance scores are involved to imply the preferences of the user observed from the domain data.

\subsection{Knowledge Customization}
We extract the above CF information from the whole interaction log. The item-item pairs and user-item pairs with the highest relevance score could provide the most confident domain knowledge about behavior preferences. However, appending the same globally confident patterns for all samples may fail to achieve the best results because they are not designed for specific samples to provide the most necessary knowledge. We propose to customize the appended domain knowledge for each sample, based on two different references:

$\bullet$ \textbf{History based.} We only provide the domain knowledge relevant to the interaction history $H^u$. For Item-2-Item information, we append the top-$k$ relevant items of each historical item, from which the LLM can infer what kind of items will be preferred by the user in this domain to rank the candidates. For User-2-Item information, we present a batch of items that are most relevant to the user characterized by his whole interaction history.

$\bullet$ \textbf{History \& Candidate based.} Since we define the recommendation task for the LLM as ranking a given set of candidate items, the domain knowledge based on the history but not relevant to the given candidates could still provide a few references for ranking these specified items. As a result, we attempt to directly indicate the relevance between the user history and the candidate items mined from the domain knowledge. For Item-2-Item, we show the top-$k$ candidates that are most frequently co-clicked with each historical item. For User-2-Item, we show which candidates are most relevant to the user.

\subsection{Knowledge Expression}\label{subsec:knowledge_expression}
With the two types of domain knowledge extracted, a natural problem is how to prompt the LLM with the knowledge in an easily understandable way. And it would be more effective to make LLM aware of the rationale hidden behind some obscure knowledge. The item attributes are already in the form of natural language, and it is necessary to translate and interpret the CF information. We discuss two ways:

$\bullet$ \textbf{Text Template.} The simplest way is applying a heuristic template to express the mined CF information in texts. For example, a frequently co-clicked item pair $(i_1, i_2)$ can be expressed as `\emph{in this domain, many users who clicked the item $i_1$ are very likely to click the item $i_2$}'. 

$\bullet$ \textbf{Reasoning Path on KG.} Though the above template involves basic conclusions, we observe that it lacks a description of the rationale, i.e. the reason why $i_1$ would be frequently co-clicked with $i_2$. The LLM needs to apply its internal knowledge to understand their correlation. If the rationale behind the relevance is directly explained, it could indicate a more essential recommendation pattern in this domain beyond a single item pair. This seems to be more efficient and accurate to introduce domain knowledge. To capture such rationales, we select the format of reasoning paths on a knowledge graph (KG), which has been widely used in explainable recommendations to interpret the association between two items~\cite{xian2019kgexplainable,wang2019kgexplainable}. Thus, we mainly apply this format to decipher the Item-2-Item information. Taking a relevant pair of movies \emph{(Zootopia, Moana)} as an example, their relevance can be interpreted by a reasoning path: \emph{Zootopia} $\stackrel{\text{has genre}}\longrightarrow$ (Animation, Adventure) $\stackrel{\text{genre of}}\longrightarrow$ \emph{Moana}. This path not only explains the correlation between the two movies, but also efficiently introduces the knowledge that making recommendations based on genres is reasonable in this domain.

We modify an existing KG-based recommendation method KPRN \cite{xian2019kgexplainable} to extract the reasoning path between two given items. First, we link all items to an external knowledge graph. For a pair of items $(i_1, i_2)$, we construct a set of paths connecting them, i.e. $P(i_1,i_2)=\{p_1,p_2, \ldots, p_T\}$. Each path consists of the two items, a sequence of entities and relations: $p=[e_1 \stackrel{r_1}\rightarrow e_2 \stackrel{r_2}\rightarrow \ldots \stackrel{r_{l}}\rightarrow e_l]$, where $e_1=i_1$ and $e_l=i_2$. Then, the relevance between the two items is defined as:
\begin{align}\label{eq:kprn}
    rel(i_1,i_2) = f(i_1,i_2|P(i_1,i_2)) = \sigma (\frac{1}{T}\sum_{t=1}^{T}s_t),\nonumber
\end{align}
where $f$ is the function to process knowledge paths and $s_t$ is the prediction score for the path $p_t$. Treating item pairs with relevance score computed by Eq.(\ref{eq:item_rel}) higher than $\theta$ as positive samples $O^+ = \{(i,j)|rel(i,j) > \theta\}$ and the others as negative samples $O^- = \{(i,j')|rel(i,j') \leq \theta\}$, we learn $f(\cdot)$ by optimizing binary classification loss. Finally, we extract the reasoning path with the highest score as the rationale of each item pair. More details about the function $f$ and the optimization can be found in~\cite{xian2019kgexplainable}.

We illustrate the integrated domain knowledge for recommendation in Figure~\ref{fig:framework}. Detailed prompts used in experiments can be found in Appendix~\ref{sec:prompt_appendix}.

\begin{table}[]
    \centering
    \caption{Statistics of the three datasets.}
    \resizebox{1.0\linewidth}{!}{
        \begin{tabular}{c|ccccc}
            \toprule
            Dataset & \#Users & \#Items & \#Interactions & Sparsity \\ 
            \midrule
            ML-1M & 6,034 & 3,533 & 575,272 & 95.35\% \\ 
            Beauty & 22,363 & 12,101 & 198,502 & 99.93\% \\ 
            Online Retail & 16,517 & 3,466 & 514,694 & 99.10\% \\
            \bottomrule
        \end{tabular}
    }
    \label{tab:statistics}
\end{table}

\begin{table*}[!ht]
 \center
  \caption{Overall recommendation performance on datasets with randomly sampled candidate items. N@k denotes NDCG@k. The best results among full training methods and zero-shot methods are both shown in bold. ``*'' denotes our paradigm significantly improves original LLMs by introducing domain knowledge, with t-test at p<0.05 level.}
  \resizebox{1.0\linewidth}{!}{
      \begin{tabular}{p{0.07\textwidth}p{0.14\textwidth}p{0.06\textwidth}p{0.06\textwidth}p{0.06\textwidth}p{0.06\textwidth}|p{0.06\textwidth}p{0.06\textwidth}p{0.06\textwidth}p{0.06\textwidth}|p{0.06\textwidth}p{0.06\textwidth}p{0.06\textwidth}p{0.06\textwidth}}
      \toprule
      \multirow{2}{*}{Model} & \multicolumn{4}{c}{ML1M} & \multicolumn{4}{c}{Beauty} & \multicolumn{4}{c}{Online Retail} \\ \cmidrule(lr){2-14} 
       & & N@1 & N@5 & N@10 & HR@10 & N@1 & N@5 & N@10 & HR@10 & N@1 & N@5 & N@10 & HR@10 \\
      \midrule
      \multirow{11}{*}{Full Train} & \multicolumn{4}{l}{Classical Recommendation Methods} \\ \cmidrule(lr){2-14} 
        & Pop & 0.2950 & 0.4881 & 0.5540 & 0.8650  & 0.1450 & 0.2862 & 0.3648 & 0.6600  & 0.2700 & 0.4864 & 0.5550 & 0.8950 \\
        & Item-based CF & 0.3518 & 0.5789 & 0.6035 & 0.8502  & 0.0950 & 0.1055 & 0.1055 & 0.1100  & 0.4300 & 0.4973 & 0.4989 & 0.5550 \\ 
        & BPR-MF & 0.3600 & 0.6115 & 0.6503 & \textbf{0.9619}  & 0.2600 & 0.3940 & 0.4456 & 0.6850  & 0.4550 & 0.6388 & 0.6900 & 0.9600 \\
        & KGAT & 0.3630  & 0.6111  & 0.6531  & 0.9600  & 0.1800 & 0.3565 & 0.4087 & 0.6800 &  - &  - &  - &  - \\
        & BERT4Rec & \textbf{0.5660}  & \textbf{0.7380}  & \textbf{0.7620}  & 0.9600  & 0.2808  & 0.4555  & 0.5102  & 0.7799  & \textbf{0.6140} & 0.7420 & 0.7780 & 0.9640 \\
        & SASRec & 0.5250 & 0.7291 & 0.7495 & 0.9550  & 0.3000 & 0.4869 & 0.5397 & 0.8100  & 0.6050  & \textbf{0.7621}  & \textbf{0.7940}  & \textbf{0.9710} \\
      \cmidrule(lr){2-14} 
      & \multicolumn{4}{l}{Fine-tuned LLM for Recommendation} \\ 
      \cmidrule(lr){2-14} 
        & P5 & 0.2550  & 0.4593  & 0.5174  & 0.8350  & 0.1230  & 0.2611  & 0.3312  & 0.6150  & 0.2670  & 0.4959  & 0.5546  & 0.8920 \\
        & Tuned Llama-2-7b & 0.5250 & 0.7033 & 0.7323 & 0.9400 & \textbf{0.4300} & \textbf{0.5878} & \textbf{0.6220} & \textbf{0.8500} & 0.6000 & 0.6939 & 0.7237 & 0.8650 \\
      \midrule
      \multirow{11}{*}{Zero-Shot} & \multicolumn{4}{l}{Zero-shot Recommendation Methods} \\ 
      \cmidrule(lr){2-14} 
       & BM25 & 0.0400 & 0.1183 & 0.1677 & 0.3500  & 0.0600 & 0.1329 & 0.2083 & 0.4400  & 0.0750  & 0.1465  & 0.1936  & 0.3700 \\
       & UniSRec & 0.1000 & 0.1970 & 0.2735 & 0.5500  & 0.1300 & 0.3317 & 0.3908 & 0.7050  & 0.1650  & 0.3091  & 0.4150  & 0.6350 \\
      \cmidrule(lr){2-14} 
      & \multicolumn{4}{l}{Prompt-based LLM for Recommendation} \\ 
      \cmidrule(lr){2-14} 
        & Llama-2-7b & 0.0300 & 0.1224 & 0.2169 & 0.5100 & 0.1000 & 0.2315 & 0.3113 & 0.6300 & 0.0700 & 0.1737 & 0.2270 & 0.4600 \\
        & Davinci & 0.1300 & 0.2922 & 0.3600 & 0.6500 & 0.1150 & 0.2300 & 0.3032 & 0.5750 & 0.1500 & 0.2431 & 0.3054 & 0.5350 \\
        & ChatGPT & 0.2750 & 0.4791 & 0.5296 & 0.8100 & 0.1700 & 0.3147 & 0.3628 & 0.6000 & 0.2550 & 0.3602 & 0.4150 & 0.6000 \\
        & ChatGPT-Recency & 0.2450 & 0.4376 & 0.5083 & 0.8300 & 0.1900 & 0.2985 & 0.3445 & 0.5450 & 0.2950 & 0.3756 & 0.4120 & 0.5700 \\
        & ChatGPT-ICL & 0.2600 & 0.4555 & 0.5046 & 0.7700 & 0.1450 & 0.2535 & 0.3152 & 0.5600 & 0.1800 & 0.3095 & 0.3657 & 0.6000 \\
        & ChatGPT-Multi & 0.1950 & 0.3040 & 0.3450 & 0.5400 & 0.1500 & 0.2909 & 0.3473 & 0.4451 & 0.1700 & 0.3018 & 0.3476 & 0.4264 \\
      \cmidrule(lr){2-14} 
        & DOKE-Davinci & 0.3300$^*$ & 0.5572$^*$ & 0.6053$^*$ & 0.9050$^*$ & 0.2200$^*$ & 0.4056$^*$ & 0.4416$^*$ & 0.6900$^*$ & 0.2750$^*$ & 0.4561$^*$ & 0.5017$^*$ & 0.7550$^*$ \\ 
        & DOKE-ChatGPT & \textbf{0.5067}$^*$ & \textbf{0.7283}$^*$ & \textbf{0.7451}$^*$ & \textbf{0.9567}$^*$ & \textbf{0.2800}$^*$ & \textbf{0.4708}$^*$ & \textbf{0.5145}$^*$ & \textbf{0.8200}$^*$ & \textbf{0.5933}$^*$ & \textbf{0.6982}$^*$ & \textbf{0.7216}$^*$ & \textbf{0.8700}$^*$ \\ 
      \bottomrule
      \end{tabular}
  }
  \label{tab:overall_performance}
\end{table*}

\section{Experiments}
We conduct empirical experiments to explore the following research questions:

\textbf{RQ1}: \emph{For the emerging paradigm of applying human-like LLMs to practical applications, will the domain knowledge introduced by our DOKE method in prompts significantly improve their performance?}


\textbf{RQ2}: \emph{Which type of domain knowledge helps to improve the performance of LLMs most?}

\textbf{RQ3}: \emph{Which samples are improved with the incorporation of domain knowledge?}

\subsection{Experimental Settings}

\subsubsection{Dataset}
We employ three public recommendation datasets for our experiments.

$\bullet$ \textbf{MovieLens-1M} (\textbf{ML1M})~\cite{harper2015movielens}: this is a popular benchmark containing 1,000,000 pieces of movie ratings, ranging from 1 to 5. For each movie, its title, release year, and genre are available.

$\bullet$ \textbf{Amazon Beauty}~\cite{ni2019amazon}: this is an e-commerce dataset about beauty products collected from Amazon. It records user interactions, including reviews and associated rating scores. The title, categories, brand, and description of each product are provided.

$\bullet$ \textbf{Online Retail}\footnote{https://www.kaggle.com/carrie1/ecommerce-data}: it contains transactions between 01/12/2010 and 09/12/2011 on a UK-based online retail platform, with only product titles available.

For all datasets, we consider users and items with at least 5 records. We first sort all interactions by the timestamps. Then, we split the last interaction of each user into the test set, the second-to-last one into the validation set, and the rest into the training data. The statistics of datasets are shown in Table~\ref{tab:statistics}.

Comparing the three datasets, information about each movie of ML1M can be available in the world knowledge used for LLMs pre-training, while the information about some products would be less accessible. Moreover, the two e-commerce datasets are sparser than ML1M. By selecting benchmarks from different domains and levels of sparsity, we expect more reliable conclusions can be drawn.

\subsubsection{Evaluation Metrics}
For each interaction in the testing set, we sample 19 random negative items and 19 hard negative items according to their click frequency to rank together respectively. We then apply the most widespread ranking metrics, including Normalized Discounted Cumulative Gain at rank k (NDCG@k) and hit ratio at rank k (HR@k).

\begin{table*}[!ht]
 \center
  \caption{Sensitivity analysis of various prompts used in our DOKE paradigm to incorporate the same domain-specific knowledge.}
  \resizebox{1.0\linewidth}{!}{
      \begin{tabular}{p{0.15\textwidth}p{0.06\textwidth}p{0.06\textwidth}p{0.06\textwidth}p{0.06\textwidth}|p{0.06\textwidth}p{0.06\textwidth}p{0.06\textwidth}p{0.06\textwidth}|p{0.06\textwidth}p{0.06\textwidth}p{0.06\textwidth}p{0.06\textwidth}}
      \toprule
      \multirow{2}{*}{Model} & \multicolumn{4}{c}{ML1M} & \multicolumn{4}{c}{Beauty} & \multicolumn{4}{c}{Online Retail} \\ \cmidrule(lr){2-13} 
        & N@1 & N@5 & N@10 & HR@10 & N@1 & N@5 & N@10 & HR@10 & N@1 & N@5 & N@10 & HR@10 \\
        \midrule
        ChatGPT & 0.2750 & 0.4791 & 0.5296 & 0.8100 & 0.1700 & 0.3147 & 0.3628 & 0.6000 & 0.2550 & 0.3602 & 0.4150 & 0.6000 \\
        \midrule
        DOKE (prompt1) & 0.5000 & 0.7199 & 0.7367 & 0.9500 & 0.2800 & 0.4751 & 0.5153 & 0.8200 & 0.5900 & 0.6918 & 0.7151 & 0.8600 \\
        DOKE (prompt2) & 0.5200 & 0.7379 & 0.7546 & 0.9600 & 0.2800 & 0.4673 & 0.5147 & 0.8200 & 0.5900 & 0.7011 & 0.7246 & 0.8800 \\
        DOKE (prompt3) & 0.5000 & 0.7272 & 0.7439 & 0.9600 & 0.2800 & 0.4699 & 0.5134 & 0.8200 & 0.6000 & 0.7018 & 0.7251 & 0.8700 \\
        \midrule
        DOKE-ChatGPT & 0.5067 & 0.7283 & 0.7451 & 0.9567 & 0.2800 & 0.4708 & 0.5145 & 0.8200 & 0.5933 & 0.6982 & 0.7216 & 0.8700 \\
      \bottomrule
      \end{tabular}
  }
  \label{tab:prompt_sensitivity}
\end{table*}

\subsubsection{Baseline}
We select four types of baselines for comparison.

$\bullet$ Classical recommendation methods: these models are fully trained on the interaction data. It includes popular methods like \textbf{Pop}, collaborative filtering algorithms (\textbf{Item-based CF}~\cite{sarwar2001itemcf} and \textbf{BPR-MF}~\cite{koren2009matrixfactor}), knowledge graph-based model \textbf{KGAT~\cite{wang2019kgat}} and sequential recommendation models, such as \textbf{SASRec}~\cite{kang2018sasrec} and \textbf{BERT4Rec~\cite{sun2019bert4rec}}.

$\bullet$ Zero-shot recommendation methods: the above methods merely cater to the source domain, without any generalization like LLMs. Thus, we also consider some zero-shot approaches, including BM25 which matches text between candidates and the interaction history, and UniSRec~\cite{hou2022unisrec} (a sequential model pre-trained on diverse recommendation datasets).

$\bullet$ Fine-tuned LLMs for recommendation: fine-tuning with training data from the target domain is a common approach for adapting LLMs to specific tasks. We select \textbf{P5}~\cite{geng2022p5} which is built on T5 by fine-tuning with multiple recommendation tasks. And we also fine-tuned \textbf{Llama-2-7b}~\cite{touvron2023llama2} on the defined recommendation task for comparison. Nonetheless, it is worth noting that this strategy is resource-intensive since LLMs have massive parameters.

$\bullet$ Prompt-based LLMs for recommendation: this strategy adapts LLMs via prompts without any parameter modification, especially feasible for those LLMs with only API accessible. We consider two powerful LLMs, i.e. \textbf{Davinci} (short for text-davinci-003) and \textbf{ChatGPT} with naive task instruction. Besides, three kinds of prompts with in-context learning or CoT techniques are included. \textbf{ChatGPT-Recency}~\cite{hou2023prompt_icl_example} emphasizes the most recent historical interaction. \textbf{ChatGPT-ICL}~\cite{hou2023prompt_icl_example} formulates the most recent interaction as a recommendation example in the prompt. And \textbf{ChatGPT-Multi} prompts the LLM to make recommendations by first summarizing user preferences and then ranking candidate items.

In this paper, we propose a general paradigm to incorporate domain-specific knowledge in prompts instead of merely prompt revision, denoted as \textbf{DOKE-Davinci} and \textbf{DOKE-ChatGPT}.





\subsubsection{Implementation Details}
All baseline implementations adhere to their original papers or open-sourced projects.

As for Llama-2-7b, we load the checkpoint from Hugging Face and fine-tune it using LoRA with a rank of 8. We set DeepSpeed zero stage as 2 to train the model on 8 $\times$ V100 (32G). The max sequence length is 1024, the learning rate of LoRA parameters is set as 5e-4, and the batch size is 32. 

In terms of other LLMs, we directly call OpenAI’s API text-davinci-003 and gpt-3.5-turbo to generate ranking results. The hyper-parameter temperature is set to 0 for reproduction. For each user's interaction history, we set the maximum length to 50. And the number of candidate items to be ranked is 20 for all models. Considering the cost of calling OpenAI APIs, we randomly evaluate 1,000 testing samples for the final results. More implementation details of each method are in Appendix~\ref{sec:Implement_Detail_Appendix}\footnote{We will release the source code after review at http://}.

\subsection{Overall Performance (RQ1)}
Table~\ref{tab:overall_performance} compares the overall performance on randomly sampled candidate items, and results on candidates sampled by popularities are in Appendix~\ref{sec:result_randneg_appendix}. We also analyze the sensitivity of prompts used in our method, and show the results in Table~\ref{tab:prompt_sensitivity}. Three primary observations can be derived.

First, LLMs have the potential to present promising recommendation accuracy, in addition to their advantages of possessing human-like capabilities and generalization for recommendation tasks. In Table~\ref{tab:overall_performance}, we can see that prompt-based LLMs without fine-tuning and external knowledge incorporation achieve a certain accuracy in the three recommendation domains, ChatGPT outperforming the zero-shot baselines BM25 and the pre-trained sequential recommendation model UniSRec. We analyze this stems from their abundant internal knowledge and impressive reasoning capability, rather than just term matching.
However, original LLMs still perform inferior to classical BPR-MF and SASRec models, which are fully trained on the collected interaction logs and learn rich knowledge to deal with this domain. After fine-tuning, the tuned Llama-2-7b model demonstrates significant improvements in recommendation accuracy, even surpassing the classical recommendation models. This promotion indicates not only the potential of LLMs but also the necessity of domain-specific knowledge for the recommendation task, which can be mainly captured from the domain data but lacking in original LLMs. Though the promotion also proves the effectiveness of fine-tuning for domain adaptation of LLMs, this approach is resource-intensive due to massive parameters. This inspires our cost-effective paradigm for introducing domain-specific knowledge without LLMs modification.

Second, our proposed paradigm DOKE significantly improves the recommendation performance of LLMs by introducing domain-specific knowledge through prompts, with a t-test at p<0.05 level. This method does not require any model fine-tuning, reducing the computational cost.
From Table~\ref{tab:overall_performance}, we find that the recommendation accuracy of Davinci and ChatGPT under our DOKE paradigm is much better than that of the original LLMs, across all datasets and all evaluation metrics. The accuracy can even be comparable to that of some classical recommendation methods and the fine-tuned Llama-2-7b. This indicates that domain knowledge is critical for practical applications and our method effectively supplements LLMs with necessary domain knowledge to enhance their performance.
In addition, since our paradigm does not modify any LLMs' parameters, it would not impact any other capabilities of LLMs which are also desirable in recommender systems and other applications.

Third, the improvements under our DOKE paradigm can be attributed to the introduction of domain-specific knowledge rather than prompt designing. And it is not sensitive to subtle modification of the prompt. 
We make a comparison with several methods that design different styles of prompts to enhance the recommendation task, i.e. ChatGPT-Recency, ChatGPT-ICL, and ChatGPT-Multi. We observe that the improvements of DOKE-ChatGPT are significantly larger than those attained by the other three approaches. This result confirms the essential effects of domain-specific knowledge introduced by our method, beyond the impacts of prompt design. In addition, we also consider several different prompts within our paradigm to include the same domain knowledge. The results shown in Table~\ref{tab:prompt_sensitivity} reaffirm that our approach augments LLMs mainly based on the introduced knowledge and demonstrates a robust insensitivity to prompt modification.

In conclusion, we answer RQ1: \textit{Domain-specific knowledge is critical to improve the performance of LLMs on practical tasks, and our proposed paradigm is effective and efficient to introduce such knowledge, without any fine-tuning cost.}

\begin{table*}[!ht]  
 \center
 \caption{Ablation studies of various types of domain knowledge on datasets with randomly sampled candidate items. `I2I' denotes Item-2-Item information, and `U2I' denotes User-2-Item. `His' denotes history and `Cand' means candidate items. The improvement percentages are calculated based on ChatGPT.}
  \resizebox{1.0\linewidth}{!}{
      \begin{tabular}{p{0.16\textwidth}ll|ll|ll}  
      \toprule  
      & \multicolumn{2}{c}{ML1M} & \multicolumn{2}{c}{Beauty} & \multicolumn{2}{c}{Online Retail} \\ \cmidrule(lr){2-7}   
        & \multicolumn{1}{c}{NDCG@1} & \multicolumn{1}{c|}{NDCG@10} & \multicolumn{1}{c}{NDCG@1} & \multicolumn{1}{c|}{NDCG@10} & \multicolumn{1}{c}{NDCG@1} & \multicolumn{1}{c}{NDCG@10} \\  
      \midrule  
        ChatGPT & 0.2750 & 0.5296 & 0.1700 & 0.3628 & 0.2550 & 0.4150 \\  
      \midrule  
        \;+ Item Attr & 0.2700~~(-1.8\%) & 0.5321~~(+0.5\%) & 0.1850~~(+8.8\%) & 0.3576~~(-1.4\%) & -- & -- \\  
      \midrule  
        \;+ global I2I CF & 0.2200~~(-20.0\%) & 0.4655~~(-12.1\%) & 0.2000~~(+17.6\%) & 0.3672~~(+1.2\%) & 0.2350~~(-7.8\%) & 0.3552~~(-14.4\%) \\  
        \;+ His I2I CF & 0.3150~~(+14.5\%) & 0.5453~~(+3.0\%) & 0.2300~~(+35.3\%) & 0.3721~~(+2.6\%) & 0.3750~~(+47.1\%) & 0.5048~~(+21.6\%) \\  
        \;+ His-Cand I2I CF & 0.3100~~(+12.7\%) & 0.5811~~(+9.7\%) & 0.2200~~(+29.4\%) & 0.4653~~(+28.3\%) & 0.2800~~(+9.8\%) & 0.5372~~(+29.4\%) \\  
      \midrule  
        \;+ His U2I CF & 0.3200~~(+16.4\%) & 0.5000~~(-5.6\%) & 0.2350~~(+38.2\%) & 0.3974~~(+9.5\%) & 0.3800~~(+49.0\%) & 0.4777~~(+15.1\%) \\  
        \;+ His-Cand U2I CF & 0.5067~~(+84.3\%) & 0.7439~~(+40.5\%) & 0.2800~~(+64.7\%) & 0.5145~~(+41.8\%) & 0.5933~~(+132.7\%) & 0.7216~~(+73.9\%) \\  
      \midrule  
        \;+ His I2I CF \& Path & 0.3100~~(+12.7\%) & 0.5503~~(+3.9\%) & 0.2400~~(+41.2\%) & 0.3924~~(+8.2\%) & -- & -- \\  
      \bottomrule  
      \end{tabular}  
  }  
  \label{tab:ablation_study_randneg}  
\end{table*}  

\subsection{Ablation Studies (RQ2)}
We conduct ablation studies to analyze which domain knowledge improves LLMs the most. The verification corresponds to the three steps of our paradigm, as shown in Table~\ref{tab:ablation_study_randneg}.

$\bullet$ \textbf{Knowledge Preparation}. The item attributes and CF information, i.e. Item-2-Item (I2I) and User-2-Item (U2I) pairs, are introduced into the LLM respectively. We observe that both types of information enhance the results. Specifically, the improvements brought by item attributes are a little slight. This may be due to that the LLM already has enough knowledge about the movies in ML-1M, and the titles of products are also informative. Therefore, introducing a few more attributes results in limited promotion. Whereas, CF signals yield significant enhancements across the three datasets, especially the U2I information. These are widely used domain knowledge captured by existing recommendation methods, while exactly lacking in the LLM. Such improvement demonstrates the importance of domain-specific knowledge and confirms the effectiveness of our proposed DOKE paradigm for incorporation.

$\bullet$ \textbf{Knowledge Customization.} Comparing the global CF information, those related to the history and those involving both history and candidates, we find that global CF information results in the least even negative improvement, while those based on both history and candidates improve the most. Though the globally relevant item pairs reflect the most confident domain knowledge, they may not be useful for the current sample. Whereas, customized knowledge can design reference information for the current sample and lead to better results. This inspires us that it is more effective to incorporate information that is relevant to the current sample.

$\bullet$ \textbf{Knowledge Expression}. We compare different expression formats of the I2I information, including text templates (i.e. His I2I) and reasoning paths on KG (i.e. His I2I + Path). This experiment is missing in Online Retail since it is hard to link its products without attributes to entities. We observe that reasoning paths on KG with explicit explanations perform slightly better than only templated texts. We analyze two reasons for the slight improvements. First, the LLM possesses rich knowledge so that it may understand most of the domain knowledge contained in the provided item pairs even without explicit explanations. We also emphasize the instruction of "understanding the provided domain knowledge" in the prompt. Second, our reasoning paths extracted from open-sourced KG may be of low quality since no specific knowledge graphs are available for these datasets. 
We speculate that it can improve more if the provided domain knowledge can be expressed in a more accurate and understandable way. 

Thus, we answer RQ2: \emph{CF information customized for each sample improves the performance of LLMs most}.

\begin{figure}
    \centering
    \includegraphics[width=0.99\linewidth]{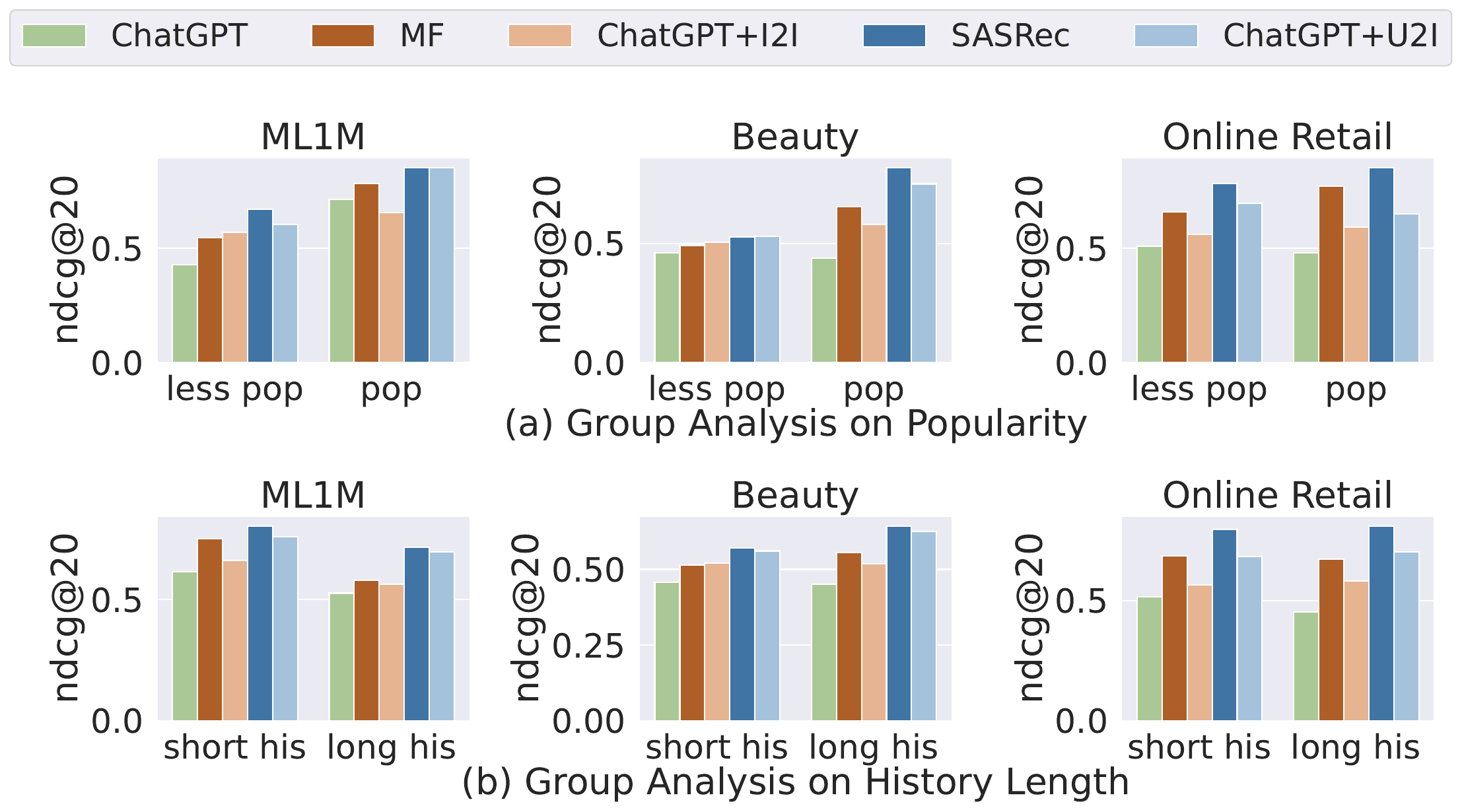}
    \caption{The results of group analysis.}
    \label{fig:group_result}
\end{figure}

\subsection{Group Analysis (RQ3)}
To analyze which samples can be improved more by incorporating domain knowledge, we divide the testing samples into two groups based on their click frequencies and the user history length respectively. We evaluate ChatGPT, ChatGPT + His-Cand I2I/U2I and the baselines (MF, SASRec) that help to extract domain knowledge on these groups. Figure~\ref{fig:group_result} illustrates the results.

In general, we observe that the incorporation of domain knowledge improves the results on samples where the baseline can achieve great accuracy by learning from the domain data. And most baselines tend to show better results on the popular group and long history group.
\emph{In these groups, more interaction records and sufficient user history for analyzing preferences are available, thus more confidential domain knowledge about recommendation patterns can be captured to improve the performance of LLMs (RQ3)}. As for groups with sparse data, such as less popular groups and short-history groups on Beauty, it is a challenge for full training models but LLMs can leverage internal knowledge to achieve better results.



\section{Conclusion}
In this paper, we demonstrate that LLMs lack domain-specific knowledge when applied to practical applications and propose a general paradigm for augmentation without resource-intensive LLMs fine-tuning, namely DOKE. It relies on a domain knowledge extractor for knowledge preparation, customization, and expression. To illustrate its efficacy, we implement the DOKE paradigm in recommender systems, incorporating item attributes and CF information. Extensive experiments verify that DOKE is both effective and efficient to introduce professional domain knowledge and significantly enhance the performance of LLMs.

Although our DOKE paradigm brings remarkable improvements, several aspects need further exploration. One promising direction involves extending the general paradigm to diverse domains. Given the diversity of data types and tasks across different domains, some case-by-case designs may be required. Furthermore, understanding how the incorporated domain knowledge influences the reasoning process within LLMs is another research issue.

\balance
\bibliographystyle{ACM-Reference-Format}
\bibliography{main}


\begin{thebibliography}{58}


\ifx \showCODEN    \undefined \def \showCODEN     #1{\unskip}     \fi
\ifx \showDOI      \undefined \def \showDOI       #1{#1}\fi
\ifx \showISBNx    \undefined \def \showISBNx     #1{\unskip}     \fi
\ifx \showISBNxiii \undefined \def \showISBNxiii  #1{\unskip}     \fi
\ifx \showISSN     \undefined \def \showISSN      #1{\unskip}     \fi
\ifx \showLCCN     \undefined \def \showLCCN      #1{\unskip}     \fi
\ifx \shownote     \undefined \def \shownote      #1{#1}          \fi
\ifx \showarticletitle \undefined \def \showarticletitle #1{#1}   \fi
\ifx \showURL      \undefined \def \showURL       {\relax}        \fi
\providecommand\bibfield[2]{#2}
\providecommand\bibinfo[2]{#2}
\providecommand\natexlab[1]{#1}
\providecommand\showeprint[2][]{arXiv:#2}

\bibitem[Bao et~al\mbox{.}(2023)]%
        {bao2023tallrec}
\bibfield{author}{\bibinfo{person}{Keqin Bao}, \bibinfo{person}{Jizhi Zhang}, \bibinfo{person}{Yang Zhang}, \bibinfo{person}{Wenjie Wang}, \bibinfo{person}{Fuli Feng}, {and} \bibinfo{person}{Xiangnan He}.} \bibinfo{year}{2023}\natexlab{}.
\newblock \showarticletitle{TALLRec: An Effective and Efficient Tuning Framework to Align Large Language Model with Recommendation}.
\newblock \bibinfo{journal}{\emph{arXiv preprint arXiv:2305.00447}} (\bibinfo{year}{2023}).
\newblock


\bibitem[Brown et~al\mbox{.}(2020)]%
        {brown2020language}
\bibfield{author}{\bibinfo{person}{Tom Brown}, \bibinfo{person}{Benjamin Mann}, \bibinfo{person}{Nick Ryder}, \bibinfo{person}{Melanie Subbiah}, \bibinfo{person}{Jared~D Kaplan}, \bibinfo{person}{Prafulla Dhariwal}, \bibinfo{person}{Arvind Neelakantan}, \bibinfo{person}{Pranav Shyam}, \bibinfo{person}{Girish Sastry}, \bibinfo{person}{Amanda Askell}, {et~al\mbox{.}}} \bibinfo{year}{2020}\natexlab{}.
\newblock \showarticletitle{Language models are few-shot learners}.
\newblock \bibinfo{journal}{\emph{Advances in neural information processing systems}}  \bibinfo{volume}{33} (\bibinfo{year}{2020}), \bibinfo{pages}{1877--1901}.
\newblock


\bibitem[Chowdhery et~al\mbox{.}(2022)]%
        {chowdhery2022palm}
\bibfield{author}{\bibinfo{person}{Aakanksha Chowdhery}, \bibinfo{person}{Sharan Narang}, \bibinfo{person}{Jacob Devlin}, \bibinfo{person}{Maarten Bosma}, \bibinfo{person}{Gaurav Mishra}, \bibinfo{person}{Adam Roberts}, \bibinfo{person}{Paul Barham}, \bibinfo{person}{Hyung~Won Chung}, \bibinfo{person}{Charles Sutton}, \bibinfo{person}{Sebastian Gehrmann}, {et~al\mbox{.}}} \bibinfo{year}{2022}\natexlab{}.
\newblock \showarticletitle{Palm: Scaling language modeling with pathways}.
\newblock \bibinfo{journal}{\emph{arXiv preprint arXiv:2204.02311}} (\bibinfo{year}{2022}).
\newblock


\bibitem[Cui et~al\mbox{.}(2022)]%
        {cui2022m6_rec}
\bibfield{author}{\bibinfo{person}{Zeyu Cui}, \bibinfo{person}{Jianxin Ma}, \bibinfo{person}{Chang Zhou}, \bibinfo{person}{Jingren Zhou}, {and} \bibinfo{person}{Hongxia Yang}.} \bibinfo{year}{2022}\natexlab{}.
\newblock \showarticletitle{M6-Rec: Generative Pretrained Language Models are Open-Ended Recommender Systems}.
\newblock \bibinfo{journal}{\emph{arXiv preprint arXiv:2205.08084}} (\bibinfo{year}{2022}).
\newblock


\bibitem[Das et~al\mbox{.}(2017)]%
        {das2017rec_survey}
\bibfield{author}{\bibinfo{person}{Debashis Das}, \bibinfo{person}{Laxman Sahoo}, {and} \bibinfo{person}{Sujoy Datta}.} \bibinfo{year}{2017}\natexlab{}.
\newblock \showarticletitle{A survey on recommendation system}.
\newblock \bibinfo{journal}{\emph{International Journal of Computer Applications}} \bibinfo{volume}{160}, \bibinfo{number}{7} (\bibinfo{year}{2017}).
\newblock


\bibitem[Devlin et~al\mbox{.}(2018)]%
        {devlin2018bert}
\bibfield{author}{\bibinfo{person}{Jacob Devlin}, \bibinfo{person}{Ming-Wei Chang}, \bibinfo{person}{Kenton Lee}, {and} \bibinfo{person}{Kristina Toutanova}.} \bibinfo{year}{2018}\natexlab{}.
\newblock \showarticletitle{Bert: Pre-training of deep bidirectional transformers for language understanding}.
\newblock \bibinfo{journal}{\emph{arXiv preprint arXiv:1810.04805}} (\bibinfo{year}{2018}).
\newblock


\bibitem[Du et~al\mbox{.}(2023)]%
        {du2023llm_enhance_query}
\bibfield{author}{\bibinfo{person}{Yingpeng Du}, \bibinfo{person}{Di Luo}, \bibinfo{person}{Rui Yan}, \bibinfo{person}{Hongzhi Liu}, \bibinfo{person}{Yang Song}, \bibinfo{person}{Hengshu Zhu}, {and} \bibinfo{person}{Jie Zhang}.} \bibinfo{year}{2023}\natexlab{}.
\newblock \showarticletitle{Enhancing Job Recommendation through LLM-based Generative Adversarial Networks}.
\newblock \bibinfo{journal}{\emph{arXiv preprint arXiv:2307.10747}} (\bibinfo{year}{2023}).
\newblock


\bibitem[Fan et~al\mbox{.}(2023)]%
        {fan2023llm_rec_survey}
\bibfield{author}{\bibinfo{person}{Wenqi Fan}, \bibinfo{person}{Zihuai Zhao}, \bibinfo{person}{Jiatong Li}, \bibinfo{person}{Yunqing Liu}, \bibinfo{person}{Xiaowei Mei}, \bibinfo{person}{Yiqi Wang}, \bibinfo{person}{Jiliang Tang}, {and} \bibinfo{person}{Qing Li}.} \bibinfo{year}{2023}\natexlab{}.
\newblock \showarticletitle{Recommender systems in the era of large language models (llms)}.
\newblock \bibinfo{journal}{\emph{arXiv preprint arXiv:2307.02046}} (\bibinfo{year}{2023}).
\newblock


\bibitem[Gao et~al\mbox{.}(2023)]%
        {gao2023chatrec}
\bibfield{author}{\bibinfo{person}{Yunfan Gao}, \bibinfo{person}{Tao Sheng}, \bibinfo{person}{Youlin Xiang}, \bibinfo{person}{Yun Xiong}, \bibinfo{person}{Haofen Wang}, {and} \bibinfo{person}{Jiawei Zhang}.} \bibinfo{year}{2023}\natexlab{}.
\newblock \showarticletitle{Chat-rec: Towards interactive and explainable llms-augmented recommender system}.
\newblock \bibinfo{journal}{\emph{arXiv preprint arXiv:2303.14524}} (\bibinfo{year}{2023}).
\newblock


\bibitem[Geng et~al\mbox{.}(2022)]%
        {geng2022p5}
\bibfield{author}{\bibinfo{person}{Shijie Geng}, \bibinfo{person}{Shuchang Liu}, \bibinfo{person}{Zuohui Fu}, \bibinfo{person}{Yingqiang Ge}, {and} \bibinfo{person}{Yongfeng Zhang}.} \bibinfo{year}{2022}\natexlab{}.
\newblock \showarticletitle{Recommendation as language processing (rlp): A unified pretrain, personalized prompt \& predict paradigm (p5)}. In \bibinfo{booktitle}{\emph{Proceedings of the 16th ACM Conference on Recommender Systems}}. \bibinfo{pages}{299--315}.
\newblock


\bibitem[Harper and Konstan(2015)]%
        {harper2015movielens}
\bibfield{author}{\bibinfo{person}{F~Maxwell Harper} {and} \bibinfo{person}{Joseph~A Konstan}.} \bibinfo{year}{2015}\natexlab{}.
\newblock \showarticletitle{The movielens datasets: History and context}.
\newblock \bibinfo{journal}{\emph{Acm transactions on interactive intelligent systems (tiis)}} \bibinfo{volume}{5}, \bibinfo{number}{4} (\bibinfo{year}{2015}), \bibinfo{pages}{1--19}.
\newblock


\bibitem[He et~al\mbox{.}(2017)]%
        {he2017neuralcf}
\bibfield{author}{\bibinfo{person}{Xiangnan He}, \bibinfo{person}{Lizi Liao}, \bibinfo{person}{Hanwang Zhang}, \bibinfo{person}{Liqiang Nie}, \bibinfo{person}{Xia Hu}, {and} \bibinfo{person}{Tat-Seng Chua}.} \bibinfo{year}{2017}\natexlab{}.
\newblock \showarticletitle{Neural collaborative filtering}. In \bibinfo{booktitle}{\emph{Proceedings of the 26th international conference on world wide web}}. \bibinfo{pages}{173--182}.
\newblock


\bibitem[He et~al\mbox{.}(2020)]%
        {he2020parade_finetune}
\bibfield{author}{\bibinfo{person}{Yun He}, \bibinfo{person}{Zhuoer Wang}, \bibinfo{person}{Yin Zhang}, \bibinfo{person}{Ruihong Huang}, {and} \bibinfo{person}{James Caverlee}.} \bibinfo{year}{2020}\natexlab{}.
\newblock \showarticletitle{PARADE: A new dataset for paraphrase identification requiring computer science domain knowledge}.
\newblock \bibinfo{journal}{\emph{arXiv preprint arXiv:2010.03725}} (\bibinfo{year}{2020}).
\newblock


\bibitem[Hou et~al\mbox{.}(2022a)]%
        {hou2022llm_enhance_encode}
\bibfield{author}{\bibinfo{person}{Yupeng Hou}, \bibinfo{person}{Shanlei Mu}, \bibinfo{person}{Wayne~Xin Zhao}, \bibinfo{person}{Yaliang Li}, \bibinfo{person}{Bolin Ding}, {and} \bibinfo{person}{Ji-Rong Wen}.} \bibinfo{year}{2022}\natexlab{a}.
\newblock \showarticletitle{Towards universal sequence representation learning for recommender systems}. In \bibinfo{booktitle}{\emph{Proceedings of the 28th ACM SIGKDD Conference on Knowledge Discovery and Data Mining}}. \bibinfo{pages}{585--593}.
\newblock


\bibitem[Hou et~al\mbox{.}(2022b)]%
        {hou2022unisrec}
\bibfield{author}{\bibinfo{person}{Yupeng Hou}, \bibinfo{person}{Shanlei Mu}, \bibinfo{person}{Wayne~Xin Zhao}, \bibinfo{person}{Yaliang Li}, \bibinfo{person}{Bolin Ding}, {and} \bibinfo{person}{Ji-Rong Wen}.} \bibinfo{year}{2022}\natexlab{b}.
\newblock \showarticletitle{Towards Universal Sequence Representation Learning for Recommender Systems}. In \bibinfo{booktitle}{\emph{Proceedings of the 28th ACM SIGKDD Conference on Knowledge Discovery and Data Mining}}. \bibinfo{pages}{585--593}.
\newblock


\bibitem[Hou et~al\mbox{.}(2023a)]%
        {hou2023large_rec_eval}
\bibfield{author}{\bibinfo{person}{Yupeng Hou}, \bibinfo{person}{Junjie Zhang}, \bibinfo{person}{Zihan Lin}, \bibinfo{person}{Hongyu Lu}, \bibinfo{person}{Ruobing Xie}, \bibinfo{person}{Julian McAuley}, {and} \bibinfo{person}{Wayne~Xin Zhao}.} \bibinfo{year}{2023}\natexlab{a}.
\newblock \showarticletitle{Large language models are zero-shot rankers for recommender systems}.
\newblock \bibinfo{journal}{\emph{arXiv preprint arXiv:2305.08845}} (\bibinfo{year}{2023}).
\newblock


\bibitem[Hou et~al\mbox{.}(2023b)]%
        {hou2023prompt_icl_example}
\bibfield{author}{\bibinfo{person}{Yupeng Hou}, \bibinfo{person}{Junjie Zhang}, \bibinfo{person}{Zihan Lin}, \bibinfo{person}{Hongyu Lu}, \bibinfo{person}{Ruobing Xie}, \bibinfo{person}{Julian McAuley}, {and} \bibinfo{person}{Wayne~Xin Zhao}.} \bibinfo{year}{2023}\natexlab{b}.
\newblock \showarticletitle{Large language models are zero-shot rankers for recommender systems}.
\newblock \bibinfo{journal}{\emph{arXiv preprint arXiv:2305.08845}} (\bibinfo{year}{2023}).
\newblock


\bibitem[Izacard et~al\mbox{.}(2022)]%
        {izacard2022retrievalcorpus}
\bibfield{author}{\bibinfo{person}{Gautier Izacard}, \bibinfo{person}{Patrick Lewis}, \bibinfo{person}{Maria Lomeli}, \bibinfo{person}{Lucas Hosseini}, \bibinfo{person}{Fabio Petroni}, \bibinfo{person}{Timo Schick}, \bibinfo{person}{Jane Dwivedi-Yu}, \bibinfo{person}{Armand Joulin}, \bibinfo{person}{Sebastian Riedel}, {and} \bibinfo{person}{Edouard Grave}.} \bibinfo{year}{2022}\natexlab{}.
\newblock \showarticletitle{Few-shot learning with retrieval augmented language models}.
\newblock \bibinfo{journal}{\emph{arXiv preprint arXiv:2208.03299}} (\bibinfo{year}{2022}).
\newblock


\bibitem[Kang and McAuley(2018)]%
        {kang2018sasrec}
\bibfield{author}{\bibinfo{person}{Wang-Cheng Kang} {and} \bibinfo{person}{Julian McAuley}.} \bibinfo{year}{2018}\natexlab{}.
\newblock \showarticletitle{Self-attentive sequential recommendation}. In \bibinfo{booktitle}{\emph{2018 IEEE international conference on data mining (ICDM)}}. IEEE, \bibinfo{pages}{197--206}.
\newblock


\bibitem[Kang et~al\mbox{.}(2023)]%
        {kang2023prompt_example_finetune}
\bibfield{author}{\bibinfo{person}{Wang-Cheng Kang}, \bibinfo{person}{Jianmo Ni}, \bibinfo{person}{Nikhil Mehta}, \bibinfo{person}{Maheswaran Sathiamoorthy}, \bibinfo{person}{Lichan Hong}, \bibinfo{person}{Ed Chi}, {and} \bibinfo{person}{Derek~Zhiyuan Cheng}.} \bibinfo{year}{2023}\natexlab{}.
\newblock \showarticletitle{Do LLMs Understand User Preferences? Evaluating LLMs On User Rating Prediction}.
\newblock \bibinfo{journal}{\emph{arXiv preprint arXiv:2305.06474}} (\bibinfo{year}{2023}).
\newblock


\bibitem[Kobsa(1994)]%
        {kobsa1994contentrec}
\bibfield{author}{\bibinfo{person}{Alfred Kobsa}.} \bibinfo{year}{1994}\natexlab{}.
\newblock \showarticletitle{User modeling and user-adapted interaction}. In \bibinfo{booktitle}{\emph{Conference companion on Human factors in computing systems}}. \bibinfo{pages}{415--416}.
\newblock


\bibitem[Koren et~al\mbox{.}(2009)]%
        {koren2009matrixfactor}
\bibfield{author}{\bibinfo{person}{Yehuda Koren}, \bibinfo{person}{Robert Bell}, {and} \bibinfo{person}{Chris Volinsky}.} \bibinfo{year}{2009}\natexlab{}.
\newblock \showarticletitle{Matrix factorization techniques for recommender systems}.
\newblock \bibinfo{journal}{\emph{Computer}} \bibinfo{volume}{42}, \bibinfo{number}{8} (\bibinfo{year}{2009}), \bibinfo{pages}{30--37}.
\newblock


\bibitem[Lazaridou et~al\mbox{.}(2022)]%
        {lazaridou2022retrievalinternet}
\bibfield{author}{\bibinfo{person}{Angeliki Lazaridou}, \bibinfo{person}{Elena Gribovskaya}, \bibinfo{person}{Wojciech Stokowiec}, {and} \bibinfo{person}{Nikolai Grigorev}.} \bibinfo{year}{2022}\natexlab{}.
\newblock \showarticletitle{Internet-augmented language models through few-shot prompting for open-domain question answering}.
\newblock \bibinfo{journal}{\emph{arXiv preprint arXiv:2203.05115}} (\bibinfo{year}{2022}).
\newblock


\bibitem[Li et~al\mbox{.}(2023)]%
        {li2023llm_enhance_query}
\bibfield{author}{\bibinfo{person}{Jinming Li}, \bibinfo{person}{Wentao Zhang}, \bibinfo{person}{Tian Wang}, \bibinfo{person}{Guanglei Xiong}, \bibinfo{person}{Alan Lu}, {and} \bibinfo{person}{Gerard Medioni}.} \bibinfo{year}{2023}\natexlab{}.
\newblock \showarticletitle{GPT4Rec: A generative framework for personalized recommendation and user interests interpretation}.
\newblock \bibinfo{journal}{\emph{arXiv preprint arXiv:2304.03879}} (\bibinfo{year}{2023}).
\newblock


\bibitem[Liang et~al\mbox{.}(2023)]%
        {liang2023taskmatrix}
\bibfield{author}{\bibinfo{person}{Yaobo Liang}, \bibinfo{person}{Chenfei Wu}, \bibinfo{person}{Ting Song}, \bibinfo{person}{Wenshan Wu}, \bibinfo{person}{Yan Xia}, \bibinfo{person}{Yu Liu}, \bibinfo{person}{Yang Ou}, \bibinfo{person}{Shuai Lu}, \bibinfo{person}{Lei Ji}, \bibinfo{person}{Shaoguang Mao}, {et~al\mbox{.}}} \bibinfo{year}{2023}\natexlab{}.
\newblock \showarticletitle{Taskmatrix. ai: Completing tasks by connecting foundation models with millions of apis}.
\newblock \bibinfo{journal}{\emph{arXiv preprint arXiv:2303.16434}} (\bibinfo{year}{2023}).
\newblock


\bibitem[Liu et~al\mbox{.}(2021a)]%
        {liu2021kgrec}
\bibfield{author}{\bibinfo{person}{Danyang Liu}, \bibinfo{person}{Jianxun Lian}, \bibinfo{person}{Zheng Liu}, \bibinfo{person}{Xiting Wang}, \bibinfo{person}{Guangzhong Sun}, {and} \bibinfo{person}{Xing Xie}.} \bibinfo{year}{2021}\natexlab{a}.
\newblock \showarticletitle{Reinforced anchor knowledge graph generation for news recommendation reasoning}. In \bibinfo{booktitle}{\emph{Proceedings of the 27th ACM SIGKDD Conference on Knowledge Discovery \& Data Mining}}. \bibinfo{pages}{1055--1065}.
\newblock


\bibitem[Liu et~al\mbox{.}(2021b)]%
        {liu2021retrievalllm}
\bibfield{author}{\bibinfo{person}{Jiacheng Liu}, \bibinfo{person}{Alisa Liu}, \bibinfo{person}{Ximing Lu}, \bibinfo{person}{Sean Welleck}, \bibinfo{person}{Peter West}, \bibinfo{person}{Ronan~Le Bras}, \bibinfo{person}{Yejin Choi}, {and} \bibinfo{person}{Hannaneh Hajishirzi}.} \bibinfo{year}{2021}\natexlab{b}.
\newblock \showarticletitle{Generated knowledge prompting for commonsense reasoning}.
\newblock \bibinfo{journal}{\emph{arXiv preprint arXiv:2110.08387}} (\bibinfo{year}{2021}).
\newblock


\bibitem[Liu et~al\mbox{.}(2023)]%
        {liu2023chatgpt_rec_eval}
\bibfield{author}{\bibinfo{person}{Junling Liu}, \bibinfo{person}{Chao Liu}, \bibinfo{person}{Renjie Lv}, \bibinfo{person}{Kang Zhou}, {and} \bibinfo{person}{Yan Zhang}.} \bibinfo{year}{2023}\natexlab{}.
\newblock \showarticletitle{Is ChatGPT a Good Recommender? A Preliminary Study}.
\newblock \bibinfo{journal}{\emph{arXiv preprint arXiv:2304.10149}} (\bibinfo{year}{2023}).
\newblock


\bibitem[Liu et~al\mbox{.}(2021c)]%
        {liu2021incontext}
\bibfield{author}{\bibinfo{person}{Jiachang Liu}, \bibinfo{person}{Dinghan Shen}, \bibinfo{person}{Yizhe Zhang}, \bibinfo{person}{Bill Dolan}, \bibinfo{person}{Lawrence Carin}, {and} \bibinfo{person}{Weizhu Chen}.} \bibinfo{year}{2021}\natexlab{c}.
\newblock \showarticletitle{What Makes Good In-Context Examples for GPT-$3 $?}
\newblock \bibinfo{journal}{\emph{arXiv preprint arXiv:2101.06804}} (\bibinfo{year}{2021}).
\newblock


\bibitem[Luo et~al\mbox{.}(2023)]%
        {luo2023catastrophic_forget}
\bibfield{author}{\bibinfo{person}{Yun Luo}, \bibinfo{person}{Zhen Yang}, \bibinfo{person}{Fandong Meng}, \bibinfo{person}{Yafu Li}, \bibinfo{person}{Jie Zhou}, {and} \bibinfo{person}{Yue Zhang}.} \bibinfo{year}{2023}\natexlab{}.
\newblock \showarticletitle{An empirical study of catastrophic forgetting in large language models during continual fine-tuning}.
\newblock \bibinfo{journal}{\emph{arXiv preprint arXiv:2308.08747}} (\bibinfo{year}{2023}).
\newblock


\bibitem[Lyu et~al\mbox{.}(2023)]%
        {lyu2023llm_enhance_kg}
\bibfield{author}{\bibinfo{person}{Hanjia Lyu}, \bibinfo{person}{Song Jiang}, \bibinfo{person}{Hanqing Zeng}, \bibinfo{person}{Yinglong Xia}, {and} \bibinfo{person}{Jiebo Luo}.} \bibinfo{year}{2023}\natexlab{}.
\newblock \showarticletitle{LLM-Rec: Personalized Recommendation via Prompting Large Language Models}.
\newblock \bibinfo{journal}{\emph{arXiv preprint arXiv:2307.15780}} (\bibinfo{year}{2023}).
\newblock


\bibitem[Ni et~al\mbox{.}(2019)]%
        {ni2019amazon}
\bibfield{author}{\bibinfo{person}{Jianmo Ni}, \bibinfo{person}{Jiacheng Li}, {and} \bibinfo{person}{Julian McAuley}.} \bibinfo{year}{2019}\natexlab{}.
\newblock \showarticletitle{Justifying recommendations using distantly-labeled reviews and fine-grained aspects}. In \bibinfo{booktitle}{\emph{Proceedings of the 2019 conference on empirical methods in natural language processing and the 9th international joint conference on natural language processing (EMNLP-IJCNLP)}}. \bibinfo{pages}{188--197}.
\newblock


\bibitem[Ouyang et~al\mbox{.}(2022)]%
        {ouyang2022instructgpt}
\bibfield{author}{\bibinfo{person}{Long Ouyang}, \bibinfo{person}{Jeffrey Wu}, \bibinfo{person}{Xu Jiang}, \bibinfo{person}{Diogo Almeida}, \bibinfo{person}{Carroll Wainwright}, \bibinfo{person}{Pamela Mishkin}, \bibinfo{person}{Chong Zhang}, \bibinfo{person}{Sandhini Agarwal}, \bibinfo{person}{Katarina Slama}, \bibinfo{person}{Alex Ray}, {et~al\mbox{.}}} \bibinfo{year}{2022}\natexlab{}.
\newblock \showarticletitle{Training language models to follow instructions with human feedback}.
\newblock \bibinfo{journal}{\emph{Advances in Neural Information Processing Systems}}  \bibinfo{volume}{35} (\bibinfo{year}{2022}), \bibinfo{pages}{27730--27744}.
\newblock


\bibitem[Resnick et~al\mbox{.}(1994)]%
        {resnick1994usercf}
\bibfield{author}{\bibinfo{person}{Paul Resnick}, \bibinfo{person}{Neophytos Iacovou}, \bibinfo{person}{Mitesh Suchak}, \bibinfo{person}{Peter Bergstrom}, {and} \bibinfo{person}{John Riedl}.} \bibinfo{year}{1994}\natexlab{}.
\newblock \showarticletitle{Grouplens: An open architecture for collaborative filtering of netnews}. In \bibinfo{booktitle}{\emph{Proceedings of the 1994 ACM conference on Computer supported cooperative work}}. \bibinfo{pages}{175--186}.
\newblock


\bibitem[Sarwar et~al\mbox{.}(2001)]%
        {sarwar2001itemcf}
\bibfield{author}{\bibinfo{person}{Badrul Sarwar}, \bibinfo{person}{George Karypis}, \bibinfo{person}{Joseph Konstan}, {and} \bibinfo{person}{John Riedl}.} \bibinfo{year}{2001}\natexlab{}.
\newblock \showarticletitle{Item-based collaborative filtering recommendation algorithms}. In \bibinfo{booktitle}{\emph{Proceedings of the 10th international conference on World Wide Web}}. \bibinfo{pages}{285--295}.
\newblock


\bibitem[Schick et~al\mbox{.}(2023)]%
        {schick2023toolformer}
\bibfield{author}{\bibinfo{person}{Timo Schick}, \bibinfo{person}{Jane Dwivedi-Yu}, \bibinfo{person}{Roberto Dess{\`\i}}, \bibinfo{person}{Roberta Raileanu}, \bibinfo{person}{Maria Lomeli}, \bibinfo{person}{Luke Zettlemoyer}, \bibinfo{person}{Nicola Cancedda}, {and} \bibinfo{person}{Thomas Scialom}.} \bibinfo{year}{2023}\natexlab{}.
\newblock \showarticletitle{Toolformer: Language models can teach themselves to use tools}.
\newblock \bibinfo{journal}{\emph{arXiv preprint arXiv:2302.04761}} (\bibinfo{year}{2023}).
\newblock


\bibitem[Shen et~al\mbox{.}(2023)]%
        {shen2023hugginggpt}
\bibfield{author}{\bibinfo{person}{Yongliang Shen}, \bibinfo{person}{Kaitao Song}, \bibinfo{person}{Xu Tan}, \bibinfo{person}{Dongsheng Li}, \bibinfo{person}{Weiming Lu}, {and} \bibinfo{person}{Yueting Zhuang}.} \bibinfo{year}{2023}\natexlab{}.
\newblock \showarticletitle{Hugginggpt: Solving ai tasks with chatgpt and its friends in huggingface}.
\newblock \bibinfo{journal}{\emph{arXiv preprint arXiv:2303.17580}} (\bibinfo{year}{2023}).
\newblock


\bibitem[Sun et~al\mbox{.}(2019)]%
        {sun2019bert4rec}
\bibfield{author}{\bibinfo{person}{Fei Sun}, \bibinfo{person}{Jun Liu}, \bibinfo{person}{Jian Wu}, \bibinfo{person}{Changhua Pei}, \bibinfo{person}{Xiao Lin}, \bibinfo{person}{Wenwu Ou}, {and} \bibinfo{person}{Peng Jiang}.} \bibinfo{year}{2019}\natexlab{}.
\newblock \showarticletitle{BERT4Rec: Sequential recommendation with bidirectional encoder representations from transformer}. In \bibinfo{booktitle}{\emph{Proceedings of the 28th ACM international conference on information and knowledge management}}. \bibinfo{pages}{1441--1450}.
\newblock


\bibitem[Touvron et~al\mbox{.}(2023a)]%
        {touvron2023llama}
\bibfield{author}{\bibinfo{person}{Hugo Touvron}, \bibinfo{person}{Thibaut Lavril}, \bibinfo{person}{Gautier Izacard}, \bibinfo{person}{Xavier Martinet}, \bibinfo{person}{Marie-Anne Lachaux}, \bibinfo{person}{Timoth{\'e}e Lacroix}, \bibinfo{person}{Baptiste Rozi{\`e}re}, \bibinfo{person}{Naman Goyal}, \bibinfo{person}{Eric Hambro}, \bibinfo{person}{Faisal Azhar}, {et~al\mbox{.}}} \bibinfo{year}{2023}\natexlab{a}.
\newblock \showarticletitle{Llama: Open and efficient foundation language models}.
\newblock \bibinfo{journal}{\emph{arXiv preprint arXiv:2302.13971}} (\bibinfo{year}{2023}).
\newblock


\bibitem[Touvron et~al\mbox{.}(2023b)]%
        {touvron2023llama2}
\bibfield{author}{\bibinfo{person}{Hugo Touvron}, \bibinfo{person}{Louis Martin}, \bibinfo{person}{Kevin Stone}, \bibinfo{person}{Peter Albert}, \bibinfo{person}{Amjad Almahairi}, \bibinfo{person}{Yasmine Babaei}, \bibinfo{person}{Nikolay Bashlykov}, \bibinfo{person}{Soumya Batra}, \bibinfo{person}{Prajjwal Bhargava}, \bibinfo{person}{Shruti Bhosale}, {et~al\mbox{.}}} \bibinfo{year}{2023}\natexlab{b}.
\newblock \showarticletitle{Llama 2: Open foundation and fine-tuned chat models}.
\newblock \bibinfo{journal}{\emph{arXiv preprint arXiv:2307.09288}} (\bibinfo{year}{2023}).
\newblock


\bibitem[Wang and Lim(2023)]%
        {wang2023prompt_rec}
\bibfield{author}{\bibinfo{person}{Lei Wang} {and} \bibinfo{person}{Ee-Peng Lim}.} \bibinfo{year}{2023}\natexlab{}.
\newblock \showarticletitle{Zero-Shot Next-Item Recommendation using Large Pretrained Language Models}.
\newblock \bibinfo{journal}{\emph{arXiv preprint arXiv:2304.03153}} (\bibinfo{year}{2023}).
\newblock


\bibitem[Wang et~al\mbox{.}(2019a)]%
        {wang2019kgat}
\bibfield{author}{\bibinfo{person}{Xiang Wang}, \bibinfo{person}{Xiangnan He}, \bibinfo{person}{Yixin Cao}, \bibinfo{person}{Meng Liu}, {and} \bibinfo{person}{Tat-Seng Chua}.} \bibinfo{year}{2019}\natexlab{a}.
\newblock \showarticletitle{Kgat: Knowledge graph attention network for recommendation}. In \bibinfo{booktitle}{\emph{Proceedings of the 25th ACM SIGKDD international conference on knowledge discovery \& data mining}}. \bibinfo{pages}{950--958}.
\newblock


\bibitem[Wang et~al\mbox{.}(2019b)]%
        {wang2019kgexplainable}
\bibfield{author}{\bibinfo{person}{Xiang Wang}, \bibinfo{person}{Dingxian Wang}, \bibinfo{person}{Canran Xu}, \bibinfo{person}{Xiangnan He}, \bibinfo{person}{Yixin Cao}, {and} \bibinfo{person}{Tat-Seng Chua}.} \bibinfo{year}{2019}\natexlab{b}.
\newblock \showarticletitle{Explainable reasoning over knowledge graphs for recommendation}. In \bibinfo{booktitle}{\emph{Proceedings of the AAAI conference on artificial intelligence}}, Vol.~\bibinfo{volume}{33}. \bibinfo{pages}{5329--5336}.
\newblock


\bibitem[Wei et~al\mbox{.}(2022a)]%
        {wei2022emergent}
\bibfield{author}{\bibinfo{person}{Jason Wei}, \bibinfo{person}{Yi Tay}, \bibinfo{person}{Rishi Bommasani}, \bibinfo{person}{Colin Raffel}, \bibinfo{person}{Barret Zoph}, \bibinfo{person}{Sebastian Borgeaud}, \bibinfo{person}{Dani Yogatama}, \bibinfo{person}{Maarten Bosma}, \bibinfo{person}{Denny Zhou}, \bibinfo{person}{Donald Metzler}, {et~al\mbox{.}}} \bibinfo{year}{2022}\natexlab{a}.
\newblock \showarticletitle{Emergent abilities of large language models}.
\newblock \bibinfo{journal}{\emph{arXiv preprint arXiv:2206.07682}} (\bibinfo{year}{2022}).
\newblock


\bibitem[Wei et~al\mbox{.}(2022b)]%
        {wei2022chain}
\bibfield{author}{\bibinfo{person}{Jason Wei}, \bibinfo{person}{Xuezhi Wang}, \bibinfo{person}{Dale Schuurmans}, \bibinfo{person}{Maarten Bosma}, \bibinfo{person}{Fei Xia}, \bibinfo{person}{Ed Chi}, \bibinfo{person}{Quoc~V Le}, \bibinfo{person}{Denny Zhou}, {et~al\mbox{.}}} \bibinfo{year}{2022}\natexlab{b}.
\newblock \showarticletitle{Chain-of-thought prompting elicits reasoning in large language models}.
\newblock \bibinfo{journal}{\emph{Advances in Neural Information Processing Systems}}  \bibinfo{volume}{35} (\bibinfo{year}{2022}), \bibinfo{pages}{24824--24837}.
\newblock


\bibitem[Wu et~al\mbox{.}(2021)]%
        {wu2021llm_enhance_encode}
\bibfield{author}{\bibinfo{person}{Chuhan Wu}, \bibinfo{person}{Fangzhao Wu}, \bibinfo{person}{Tao Qi}, {and} \bibinfo{person}{Yongfeng Huang}.} \bibinfo{year}{2021}\natexlab{}.
\newblock \showarticletitle{Empowering news recommendation with pre-trained language models}. In \bibinfo{booktitle}{\emph{Proceedings of the 44th International ACM SIGIR Conference on Research and Development in Information Retrieval}}. \bibinfo{pages}{1652--1656}.
\newblock


\bibitem[Wu et~al\mbox{.}(2023a)]%
        {wu2023visual_chatgpt}
\bibfield{author}{\bibinfo{person}{Chenfei Wu}, \bibinfo{person}{Shengming Yin}, \bibinfo{person}{Weizhen Qi}, \bibinfo{person}{Xiaodong Wang}, \bibinfo{person}{Zecheng Tang}, {and} \bibinfo{person}{Nan Duan}.} \bibinfo{year}{2023}\natexlab{a}.
\newblock \showarticletitle{Visual chatgpt: Talking, drawing and editing with visual foundation models}.
\newblock \bibinfo{journal}{\emph{arXiv preprint arXiv:2303.04671}} (\bibinfo{year}{2023}).
\newblock


\bibitem[Wu et~al\mbox{.}(2019)]%
        {wu2019attention}
\bibfield{author}{\bibinfo{person}{Libing Wu}, \bibinfo{person}{Cong Quan}, \bibinfo{person}{Chenliang Li}, \bibinfo{person}{Qian Wang}, \bibinfo{person}{Bolong Zheng}, {and} \bibinfo{person}{Xiangyang Luo}.} \bibinfo{year}{2019}\natexlab{}.
\newblock \showarticletitle{A context-aware user-item representation learning for item recommendation}.
\newblock \bibinfo{journal}{\emph{ACM Transactions on Information Systems (TOIS)}} \bibinfo{volume}{37}, \bibinfo{number}{2} (\bibinfo{year}{2019}), \bibinfo{pages}{1--29}.
\newblock


\bibitem[Wu et~al\mbox{.}(2023b)]%
        {wu2023llm_rec_survey}
\bibfield{author}{\bibinfo{person}{Likang Wu}, \bibinfo{person}{Zhi Zheng}, \bibinfo{person}{Zhaopeng Qiu}, \bibinfo{person}{Hao Wang}, \bibinfo{person}{Hongchao Gu}, \bibinfo{person}{Tingjia Shen}, \bibinfo{person}{Chuan Qin}, \bibinfo{person}{Chen Zhu}, \bibinfo{person}{Hengshu Zhu}, \bibinfo{person}{Qi Liu}, {et~al\mbox{.}}} \bibinfo{year}{2023}\natexlab{b}.
\newblock \showarticletitle{A Survey on Large Language Models for Recommendation}.
\newblock \bibinfo{journal}{\emph{arXiv preprint arXiv:2305.19860}} (\bibinfo{year}{2023}).
\newblock


\bibitem[Xi et~al\mbox{.}(2023)]%
        {xi2023llm_enhance_kg}
\bibfield{author}{\bibinfo{person}{Yunjia Xi}, \bibinfo{person}{Weiwen Liu}, \bibinfo{person}{Jianghao Lin}, \bibinfo{person}{Jieming Zhu}, \bibinfo{person}{Bo Chen}, \bibinfo{person}{Ruiming Tang}, \bibinfo{person}{Weinan Zhang}, \bibinfo{person}{Rui Zhang}, {and} \bibinfo{person}{Yong Yu}.} \bibinfo{year}{2023}\natexlab{}.
\newblock \showarticletitle{Towards Open-World Recommendation with Knowledge Augmentation from Large Language Models}.
\newblock \bibinfo{journal}{\emph{arXiv preprint arXiv:2306.10933}} (\bibinfo{year}{2023}).
\newblock


\bibitem[Xian et~al\mbox{.}(2019)]%
        {xian2019kgexplainable}
\bibfield{author}{\bibinfo{person}{Yikun Xian}, \bibinfo{person}{Zuohui Fu}, \bibinfo{person}{Shan Muthukrishnan}, \bibinfo{person}{Gerard De~Melo}, {and} \bibinfo{person}{Yongfeng Zhang}.} \bibinfo{year}{2019}\natexlab{}.
\newblock \showarticletitle{Reinforcement knowledge graph reasoning for explainable recommendation}. In \bibinfo{booktitle}{\emph{Proceedings of the 42nd international ACM SIGIR conference on research and development in information retrieval}}. \bibinfo{pages}{285--294}.
\newblock


\bibitem[Xiao et~al\mbox{.}(2022)]%
        {xiao2022llm_enhance_encode}
\bibfield{author}{\bibinfo{person}{Shitao Xiao}, \bibinfo{person}{Zheng Liu}, \bibinfo{person}{Yingxia Shao}, \bibinfo{person}{Tao Di}, \bibinfo{person}{Bhuvan Middha}, \bibinfo{person}{Fangzhao Wu}, {and} \bibinfo{person}{Xing Xie}.} \bibinfo{year}{2022}\natexlab{}.
\newblock \showarticletitle{Training large-scale news recommenders with pretrained language models in the loop}. In \bibinfo{booktitle}{\emph{Proceedings of the 28th ACM SIGKDD Conference on Knowledge Discovery and Data Mining}}. \bibinfo{pages}{4215--4225}.
\newblock


\bibitem[Yu et~al\mbox{.}(2023)]%
        {yu2023retrievalcorpus}
\bibfield{author}{\bibinfo{person}{Wenhao Yu}, \bibinfo{person}{Zhihan Zhang}, \bibinfo{person}{Zhenwen Liang}, \bibinfo{person}{Meng Jiang}, {and} \bibinfo{person}{Ashish Sabharwal}.} \bibinfo{year}{2023}\natexlab{}.
\newblock \showarticletitle{Improving Language Models via Plug-and-Play Retrieval Feedback}.
\newblock \bibinfo{journal}{\emph{arXiv preprint arXiv:2305.14002}} (\bibinfo{year}{2023}).
\newblock


\bibitem[Zeng et~al\mbox{.}(2022)]%
        {zeng2022glm}
\bibfield{author}{\bibinfo{person}{Aohan Zeng}, \bibinfo{person}{Xiao Liu}, \bibinfo{person}{Zhengxiao Du}, \bibinfo{person}{Zihan Wang}, \bibinfo{person}{Hanyu Lai}, \bibinfo{person}{Ming Ding}, \bibinfo{person}{Zhuoyi Yang}, \bibinfo{person}{Yifan Xu}, \bibinfo{person}{Wendi Zheng}, \bibinfo{person}{Xiao Xia}, {et~al\mbox{.}}} \bibinfo{year}{2022}\natexlab{}.
\newblock \showarticletitle{Glm-130b: An open bilingual pre-trained model}.
\newblock \bibinfo{journal}{\emph{arXiv preprint arXiv:2210.02414}} (\bibinfo{year}{2022}).
\newblock


\bibitem[Zhang et~al\mbox{.}(2023)]%
        {zhang2023recommendation}
\bibfield{author}{\bibinfo{person}{Junjie Zhang}, \bibinfo{person}{Ruobing Xie}, \bibinfo{person}{Yupeng Hou}, \bibinfo{person}{Wayne~Xin Zhao}, \bibinfo{person}{Leyu Lin}, {and} \bibinfo{person}{Ji-Rong Wen}.} \bibinfo{year}{2023}\natexlab{}.
\newblock \showarticletitle{Recommendation as instruction following: A large language model empowered recommendation approach}.
\newblock \bibinfo{journal}{\emph{arXiv preprint arXiv:2305.07001}} (\bibinfo{year}{2023}).
\newblock


\bibitem[Zhang et~al\mbox{.}(2022)]%
        {zhang2022opt}
\bibfield{author}{\bibinfo{person}{Susan Zhang}, \bibinfo{person}{Stephen Roller}, \bibinfo{person}{Naman Goyal}, \bibinfo{person}{Mikel Artetxe}, \bibinfo{person}{Moya Chen}, \bibinfo{person}{Shuohui Chen}, \bibinfo{person}{Christopher Dewan}, \bibinfo{person}{Mona Diab}, \bibinfo{person}{Xian Li}, \bibinfo{person}{Xi~Victoria Lin}, {et~al\mbox{.}}} \bibinfo{year}{2022}\natexlab{}.
\newblock \showarticletitle{Opt: Open pre-trained transformer language models}.
\newblock \bibinfo{journal}{\emph{arXiv preprint arXiv:2205.01068}} (\bibinfo{year}{2022}).
\newblock


\bibitem[Zhang et~al\mbox{.}(2021)]%
        {zhang2021language_rec}
\bibfield{author}{\bibinfo{person}{Yuhui Zhang}, \bibinfo{person}{Hao Ding}, \bibinfo{person}{Zeren Shui}, \bibinfo{person}{Yifei Ma}, \bibinfo{person}{James Zou}, \bibinfo{person}{Anoop Deoras}, {and} \bibinfo{person}{Hao Wang}.} \bibinfo{year}{2021}\natexlab{}.
\newblock \showarticletitle{Language models as recommender systems: Evaluations and limitations}.
\newblock  (\bibinfo{year}{2021}).
\newblock


\bibitem[Zhao et~al\mbox{.}(2020)]%
        {zhao2020kgexplainable}
\bibfield{author}{\bibinfo{person}{Kangzhi Zhao}, \bibinfo{person}{Xiting Wang}, \bibinfo{person}{Yuren Zhang}, \bibinfo{person}{Li Zhao}, \bibinfo{person}{Zheng Liu}, \bibinfo{person}{Chunxiao Xing}, {and} \bibinfo{person}{Xing Xie}.} \bibinfo{year}{2020}\natexlab{}.
\newblock \showarticletitle{Leveraging demonstrations for reinforcement recommendation reasoning over knowledge graphs}. In \bibinfo{booktitle}{\emph{Proceedings of the 43rd international ACM SIGIR conference on research and development in information retrieval}}. \bibinfo{pages}{239--248}.
\newblock


\end{thebibliography}
\appendix
\clearpage


\section{Implementation Details} \label{sec:Implement_Detail_Appendix}
We follow their original papers or open-sourced projects to implement all baselines. And we call OpenAI's APIs text-davinci-003 and gpt-3.5-turbo to generate ranking results. In the following, we will introduce how we derive different types of domain knowledge.

\textbf{Item Attributes}. We incorporate the genre and the published year of each movie on ML1M, the brand and category of each product on Beauty. Online Retail is not considered due to it having only available product titles. We directly append these attributes after each item's title.

\textbf{Item-2-Item CF}. We calculate the relevance score of all item pairs of the dataset using Eq.(\ref{eq:item_rel}). For global I2I CF, we select 20 item pairs with the globally highest relevance score as the domain knowledge. For history Item-2-Item CF, we only pay attention to the items that are most relevant to the user's historical items. Owing to the constraint of maximum prompt length, we only consider the user's 10 most recent historical items. And for each historical item, we get the top-3 relevant items. For history-candidate Item-2-Item CF, it is similar to history Item-2-Item CF, except that for each historical item, we get the top-3 relevant candidate items rather than those from the whole item set.

\textbf{User-2-Item CF}. We calculate the relevance score between each user and all items in the dataset using Eq.(\ref{eq:user_item_rel}). The representation vector of each user and item is obtained from the SASRec model trained on the corresponding dataset. For history User-2-Item CF, we select the top 20 relevant items in the whole set to imply the user's preferences. For history-candidate User-2-Item CF, we provide the user's preferences on these candidate items in a descending order inferred by the domain knowledge.

\textbf{Item-2-Item CF Path}
We modify the existing KG-based recommendation method KPRN~\cite{xian2019kgexplainable} to obtain the reasoning path between a pair of items. We use Wikidata\footnote{https://www.wikidata.org/} as the basic knowledge graph for both ML1M and Beauty. First, we build a graph based on the original dataset, considering the co-click relation between items and that between each item and the corresponding attributes. Second, we link all items in the set $\mathcal{I}$ to the Wikidata knowledge graph by calculating the TF-IDF matching score between each item title and the entities. For each item, we consider the highest-scored entity with a matching score above a threshold of 0.7 as the linked entity. We also maintain 2-order neighbors as well as the relations for each linked entity. Finally, we merge the graph on the original dataset and the incorporated part to obtain a knowledge graph on this specific domain as $\mathcal{G} = \{(h,r,t)|h,t \in \epsilon \cup \mathcal{I}, r \in R\}$, based on which we construct the connecting paths between any given pair of items, where $\epsilon$ denotes the incorporated entity set. For all item pairs, we divide those with a relevance score calculated by Eq.(\ref{eq:item_rel}) higher than $\theta = 0.2$ as positive samples $O^+$, and the others as negative samples $O^-$. Then, we optimize the model by minimizing a binary classification loss as:
\begin{align}
    \mathcal{L} = -\sum_{(i,j^+)\in O^+} \log(rel(i, j^+)) \\ \nonumber
    + \sum_{(i,j^-) \in O^-} \log (1 - rel(i, j^-)).
\end{align}
The calculation of $rel(i,j^+)$ is described in Sec~\ref{subsec:knowledge_expression}.
With the trained function $f(\cdot)$, we extract the reasoning path with the highest score as the rationale of each Item-2-Item CF pair.

\begin{table*}[!ht]
 \center
  \caption{Overall recommendation performance on datasets with candidate items sampled according to popularities. N@k denotes NDCG@k. The best results among full training methods and zero-shot methods are both shown in bold. ``*'' denotes our paradigm significantly improves original LLMs by introducing domain knowledge, with t-test at p<0.05 level.}
  \resizebox{1.0\linewidth}{!}{
      \begin{tabular}{p{0.07\textwidth}p{0.14\textwidth}p{0.06\textwidth}p{0.06\textwidth}p{0.06\textwidth}p{0.06\textwidth}|p{0.06\textwidth}p{0.06\textwidth}p{0.06\textwidth}p{0.06\textwidth}|p{0.06\textwidth}p{0.06\textwidth}p{0.06\textwidth}p{0.06\textwidth}}
      \toprule
      \multirow{2}{*}{Model} & \multicolumn{4}{c}{ML1M} & \multicolumn{4}{c}{Beauty} & \multicolumn{4}{c}{Online Retail} \\ \cmidrule(lr){2-14} 
       & & N@1 & N@5 & N@10 & HR@10 & N@1 & N@5 & N@10 & HR@10 & N@1 & N@5 & N@10 & HR@10 \\
      \midrule
      \multirow{11}{*}{Full Train} & \multicolumn{4}{l}{Classical Recommendation Methods} \\ \cmidrule(lr){2-14} 
        & Pop & 0.0500 & 0.1414 & 0.2122 & 0.4591 & 0.0200 & 0.0718 & 0.1179 & 0.2750 & 0.0850 & 0.1934 & 0.2644 & 0.5320 \\
        & Item-based CF & 0.1920 & 0.3656 & 0.4340 & 0.7480 & 0.0850 & 0.1008 & 0.1008 & 0.1100 & 0.3000 & 0.4290 & 0.4355 & 0.5180 \\
        & BPR-MF & 0.2500 & 0.4648 & 0.5340 & 0.8820 & 0.1600 & 0.3270 & 0.4089 & 0.7400 & 0.2600 & 0.4522 & 0.5097 & 0.8120 \\
        & KGAT & 0.2350 & 0.4645 & 0.5269 & 0.8900 & 0.1800 & 0.3435 & 0.4120 & 0.7210 & - & - & - & - \\
        & BERT4Rec &  0.4120 & 0.5838 & 0.6271 & 0.8704 & 0.1962 & 0.3123 & 0.3789 & 0.6254 & 0.4240 & 0.5790 & 0.6250 & 0.8590 \\
        & SASRec & \textbf{0.4580} & \textbf{0.6434} & \textbf{0.6826} & \textbf{0.9190} & 0.3250 & 0.4766 & 0.5413 & 0.8100 & \textbf{0.4950} & \textbf{0.6442} & \textbf{0.6837} & \textbf{0.9000} \\
      \cmidrule(lr){2-14} 
      & \multicolumn{4}{l}{Fine-tuned LLM for Recommendation} \\ 
      \cmidrule(lr){2-14} 
        & P5 & 0.0670 & 0.1625 & 0.2391 & 0.5040 & 0.0600 & 0.1524 & 0.2234 & 0.4730 & 0.0710 & 0.1661 & 0.2493 & 0.5240 \\
        & Tuned Llama-2-7b & 0.4250 & 0.5692 & 0.6152 & 0.8500 & \textbf{0.4300} & \textbf{0.5878} & \textbf{0.6220} & \textbf{0.8500} & 0.4300 & 0.5473 & 0.5894 & 0.7800 \\
      \midrule
      \multirow{11}{*}{Zero-Shot} & \multicolumn{4}{l}{Zero-shot Recommendation Methods} \\ 
      \cmidrule(lr){2-14} 
        & BM25 & 0.0500 & 0.1622 & 0.2373 & 0.8450 & 0.0500 & 0.1425 & 0.2104 & 0.4500 & 0.0700 & 0.1892 & 0.2410 & 0.5650 \\
        & UniSRec & 0.0600 & 0.1803 & 0.2625 & 0.6200 & 0.1250 & 0.3122 & 0.3848 & 0.6900 & 0.1500 & 0.2703 & 0.3416 & 0.6100 \\
      \cmidrule(lr){2-14} 
      & \multicolumn{4}{l}{Prompt-based LLM for Recommendation} \\ 
      \cmidrule(lr){2-14} 
        & Llama-2-7b & 0.0550 & 0.1476 & 0.2193 & 0.4650 & 0.0800 & 0.2067 & 0.2847 & 0.5950 & 0.0500 & 0.1566 & 0.2116 & 0.4350 \\
        & Davinci & 0.0800 & 0.1746 & 0.2502 & 0.5200 & 0.0900 & 0.2427 & 0.3243 & 0.6350 & 0.1400 & 0.2431 & 0.3033 & 0.5350 \\
        & ChatGPT & 0.0600 & 0.1571 & 0.2313 & 0.4900 & 0.1600 & 0.2691 & 0.3303 & 0.5600 & 0.1550 & 0.2623 & 0.3215 & 0.5400 \\
        & ChatGPT-Recency & 0.0750 & 0.1618 & 0.2257 & 0.4550 & 0.1400 & 0.2659 & 0.3249 & 0.5600 & 0.1900 & 0.2813 & 0.3336 & 0.5200 \\ 
        & ChatGPT-ICL & 0.0900 & 0.2023 & 0.2769 & 0.5550 & 0.1350 & 0.2568 & 0.3197 & 0.5700 & 0.2050 & 0.3196 & 0.3743 & 0.5900 \\
        & ChatGPT-Multi & 0.1000 & 0.2091 & 0.2910 & 0.5700 & 0.1450 & 0.2760 & 0.3326 & 0.5750 & 0.1650 & 0.2778 & 0.3275 & 0.5350 \\
      \cmidrule(lr){2-14} 
        & DOKE-Davinci & 0.1250$^*$ & 0.2932$^*$ & 0.3785$^*$ & 0.7300$^*$ & 0.1650$^*$ & 0.3080$^*$ & 0.3660$^*$ & 0.6300$^*$ & 0.1900$^*$ & 0.3569$^*$ & 0.4272$^*$ & 0.7300$^*$ \\ 
        & DOKE-ChatGPT & \textbf{0.4183}$^*$ & \textbf{0.6030}$^*$ & \textbf{0.6493}$^*$ & \textbf{0.8933}$^*$ & \textbf{0.3300}$^*$ & \textbf{0.4809}$^*$ & \textbf{0.5415}$^*$ & \textbf{0.8017}$^*$ & \textbf{0.4750}$^*$ & \textbf{0.6183}$^*$ & \textbf{0.6567}$^*$ & \textbf{0.8650}$^*$ \\ 
      \bottomrule
      \end{tabular}
  }
  \label{tab:overall_performance_pop}
\end{table*}

\begin{table*}[!ht]
 \center
  \caption{Sensitivity analysis of various prompts used in our DOKE paradigm to incorporate the same domain-specific knowledge.}
  \resizebox{1.0\linewidth}{!}{
      \begin{tabular}{p{0.15\textwidth}p{0.06\textwidth}p{0.06\textwidth}p{0.06\textwidth}p{0.06\textwidth}|p{0.06\textwidth}p{0.06\textwidth}p{0.06\textwidth}p{0.06\textwidth}|p{0.06\textwidth}p{0.06\textwidth}p{0.06\textwidth}p{0.06\textwidth}}
      \toprule
      \multirow{2}{*}{Model} & \multicolumn{4}{c}{ML1M} & \multicolumn{4}{c}{Beauty} & \multicolumn{4}{c}{Online Retail} \\ \cmidrule(lr){2-13} 
        & N@1 & N@5 & N@10 & HR@10 & N@1 & N@5 & N@10 & HR@10 & N@1 & N@5 & N@10 & HR@10 \\
        \midrule
        ChatGPT & 0.0600 & 0.1571 & 0.2313 & 0.4900 & 0.1600 & 0.2691 & 0.3303 & 0.5600 & 0.1550 & 0.2623 & 0.3215 & 0.5400 \\
        \midrule
        DOKE-prompt1 & 0.4400 & 0.6305 & 0.6730 & 0.9150 & 0.3300 & 0.4835 & 0.5434 & 0.8050 & 0.4700 & 0.6165 & 0.6560 & 0.8700 \\
        DOKE-prompt2 & 0.4000 & 0.5773 & 0.6264 & 0.8700 & 0.3350 & 0.4803 & 0.5423 & 0.8000 & 0.4750 & 0.6101 & 0.6481 & 0.8450 \\
        DOKE-prompt3 & 0.4150 & 0.6011 & 0.6484 & 0.8950 & 0.3250 & 0.4789 & 0.5388 & 0.8000 & 0.4800 & 0.6282 & 0.6661 & 0.8800 \\
        \midrule
        DOKE-ChatGPT & 0.4183 & 0.6030 & 0.6493 & 0.8933 & 0.3300 & 0.4809 & 0.5415 & 0.8017 & 0.4750 & 0.6183 & 0.6567 & 0.8650 \\
      \bottomrule
      \end{tabular}
  }
  \label{tab:prompt_sensitivity_pop}
\end{table*}

\begin{table*}[!ht]  
 \center  
 \caption{Ablation studies of various types of domain knowledge on candidate items sampled by popularities. `I2I' denotes Item-2-Item, and `U2I' denotes User-2-Item. The percentages are calculated based on ChatGPT.}  
  \resizebox{1.0\linewidth}{!}{  
      \begin{tabular}{p{0.16\textwidth}ll|ll|ll}  
      \toprule  
      & \multicolumn{2}{c}{ML1M} & \multicolumn{2}{c}{Beauty} & \multicolumn{2}{c}{Online Retail} \\ \cmidrule(lr){2-7}   
        & \multicolumn{1}{c}{NDCG@1} & \multicolumn{1}{c|}{NDCG@10} & \multicolumn{1}{c}{NDCG@1} & \multicolumn{1}{c|}{NDCG@10} & \multicolumn{1}{c}{NDCG@1} & \multicolumn{1}{c}{NDCG@10} \\  
      \midrule  
        ChatGPT & 0.0600 & 0.2313 & 0.1600 & 0.3303 & 0.1550 & 0.3215 \\  
      \midrule  
        \;+ Item Attr & 0.0750~~(+25.0\%) & 0.2632~~(+13.8\%) & 0.1650~~(+3.1\%) & 0.3368~~(+2.0\%) &  - &  -  \\  
      \midrule  
        \;+ global I2I CF & 0.0700~~(+16.7\%) & 0.2325~~(+0.5\%) & 0.1550~~(-3.1\%) & 0.3340~~(+1.1\%) & 0.1750~~(+12.9\%) & 0.3293~~(+2.4\%) \\  
        \;+ His I2I CF & 0.1600~~(+166.7\%) & 0.3269~~(+41.3\%) & 0.2250~~(40.6\%) & 0.3758~~(+13.8\%) & 0.2900~~(+87.1\%) & 0.4297~~(+33.7\%) \\  
        \;+ His-Cand I2I CF & 0.1700~~(+183.3\%) & 0.4819~~(+108.3\%) & 0.1600~~(+0.0\%) & 0.3875~~(+17.3\%) & 0.2350~~(+51.6\%) & 0.4928~~(+53.3\%) \\  
      \midrule  
        \;+ His U2I CF & 0.1950~~(+225.0\%) & 0.3581~~(+54.8\%) & 0.2400~~(+50.0\%) & 0.3957~~(+19.8\%) & 0.3000~~(+93.5\%) & 0.4018~~(+25.0\%) \\  
        \;+ His-Cand U2I CF & 0.4183~~(+597.2\%) & 0.6493~~(+180.7\%)	& 0.3300~~(+106.3\%) & 0.5415~~(+63.9\%) & 0.4750~~(+206.5\%) & 0.6567~~(+104.3\%) \\
        
      \midrule  
        \;+ His I2I CF \& Path & 0.1500~~(+150.0\%) & 0.3303~~(+42.8\%) & 0.2250~~(+40.6\%) & 0.3961~~(+19.9\%) &  - &  - \\   
      \bottomrule  
      \end{tabular}  
  }  
  \label{tab:ablation_study_pop}  
\end{table*}  

\section{Results on Popularity-based Candidates} \label{sec:result_randneg_appendix}
The overall performance and results of ablation studies on candidate items sampled by their click frequency are illustrated in Table~\ref{tab:overall_performance_pop}, Table~\ref{tab:prompt_sensitivity_pop} and Table~\ref{tab:ablation_study_pop} respectively. Conclusions that are similar to the dataset with candidate items sampled by popularities can be drawn from these results.

\section{Prompts for ML1M Dataset} \label{sec:prompt_appendix}
\subsection{Prompt template}
\noindent\textbf{Method: LLM without domain knowledge} \\
{\ttfamily\small
    Prompt template: You are a movie recommender system now. \\
    Here is the watching history of a user in the past in order:
    \meta{history titles} \\
    Now there are 20 candidate movies that this user can watch next:
    \meta{candidate titles} \\
    Please rank the 20 candidate movies by measuring the possibilities that this user would like to watch next most, according to the provided watching history. Please think step by step. \\
    Your output is only allowed to be a rerank of the candidate list. Do not add movies out of the candidate list. \\
    Please give the results as JSON array (from highest to lowest priority, movie names only): 
}

\vspace{5mm}
\noindent\textbf{Method: LLM + Item Attr} \\
{\ttfamily\small
    Prompt template: You are a movie recommender system now. \\
    Here is the watching history of a user in the past in order, \textbf{along with their features in bracket after titles}:
    \meta{history titles with feature} \\
    Now there are 20 candidate movies that this user can watch next, \textbf{also with their features}:
    \meta{candidate titles with feature} \\
    Please rank the 20 candidate movies by measuring the possibilities that this user would like to watch next most, according to the provided watching history. Please think step by step. \\
    Your output is only allowed to be a rerank of the candidate list. Do not add movies out of the candidate list. \\
    Please give the results as JSON array (from highest to lowest priority, movie names only): 
}

\vspace{5mm}
\noindent\textbf{Method: LLM + global I2I CF} \\
{\ttfamily\small
    Prompt template: You are a movie recommender system now. \\
    Here is the watching history of a user in the past in order:
    \meta{history titles} \\
    \textbf{To help you generate better recommendation, I will provide you with some collaborative filtering information. I will give you the most freqyently co-watched movie pairs on this platform.} \\
    \meta{global I2I text} \\
    Now there are 20 candidate movies that this user can watch next:
    \meta{candidate titles} \\
    Please rank the 20 candidate movies by measuring the possibilities that this user would like to watch next most, according to the provided watching history. Please think step by step. \\
    Your output is only allowed to be a rerank of the candidate list. Do not add movies out of the candidate list. \\
    Please give the results as JSON array (from highest to lowest priority, movie names only): 
}

\vspace{5mm}
\noindent\textbf{Method: LLM + His I2I CF} \\
{\ttfamily\small
    Prompt template: You are a movie recommender system now. \\
    Here is the watching history of a user in the past in order:
    \meta{history titles} \\
    To help you generate better recommendation, I will provide you with some collaborative filtering information \textbf{related to this user's historical watched movies on this platform. For each historical watched movie, I will give you its most frequently co-watched movies by all users.} \\
    \meta{His I2I text} \\
    Now there are 20 candidate movies that this user can watch next:
    \meta{candidate titles} \\
    Please rank the 20 candidate movies by measuring the possibilities that this user would like to watch next most, according to the provided watching history and the CF information. Please think step by step.\\
    Your output is only allowed to be a rerank of the candidate list. Do not add movies out of the candidate list.\\
    Please give the results as JSON array (from highest to lowest priority, movie names only):
}

\vspace{5mm}
\noindent\textbf{Method: LLM + His-Cand I2I CF} \\
{\ttfamily\small
    Prompt template: You are a movie recommender system now. \\
    Here is the watching history of a user in the past in order:
    \meta{history titles} \\
    Now there are 20 candidate movies that this user can watch next:
    \meta{candidate titles} \\
    To help you generate better recommendation, I will provide you with some collaborative filtering information related to this user's historical watched movies on this platform. For each historical watched movie, I will give you its most frequently co-watched movies \textbf{in the candidate list} by all users. \\
    \meta{His-Cand I2I text} \\
    Please rank the 20 candidate movies by measuring the possibilities that this user would like to watch next most, according to the provided watching history and the CF information. Please think step by step.\\
    Your output is only allowed to be a rerank of the candidate list. Do not add movies out of the candidate list.\\
    Please give the results as JSON array (from highest to lowest priority, movie names only):
}

\vspace{5mm}
\noindent\textbf{Method: LLM + His U2I CF} \\
{\ttfamily\small
    Prompt template: You are a movie recommender system now. \\
    Here is the watching history of a user in the past in order:
    \meta{history titles} \\
    Now there are 20 candidate movies that this user can watch next:
    \meta{candidate titles} \\
    To help you generate better recommendation, \textbf{I will give you some movies that this user may be interested in according to the history, and ranked in descending order.} They are: 
    \meta{His U2I item titles} \\
    Please rank the 20 candidate movies by measuring the possibilities that this user would like to watch next most, according to the provided watching history and the interested movies. Please think step by step. \\
    Your output is only allowed to be a rerank of the candidate list. Do not add movies out of the candidate list. \\
    Please give the results as JSON array (from highest to lowest priority, movie names only):
}

\vspace{5mm}
\noindent\textbf{Method: LLM + His-Cand U2I CF} \\
{\ttfamily\small
    Prompt template: You are a movie recommender system now. \\
    Here is the watching history of a user in the past in order:
    \meta{history titles} \\
    Now there are 20 candidate movies that this user can watch next:
    \meta{candidate titles} \\
    To help you generate better recommendation, I will give you some movies that this user may be interested in \textbf{in the candidate list} according to the history, and ranked in descending order. They are: 
    \meta{His-Cand U2I item titles} \\
    Please rank the 20 candidate movies by measuring the possibilities that this user would like to watch next most, according to the provided watching history and the interested movies. Please think step by step. \\
    Your output is only allowed to be a rerank of the candidate list. Do not add movies out of the candidate list. \\
    Please give the results as JSON array (from highest to lowest priority, movie names only):
}

\vspace{5mm}
\noindent\textbf{Method: LLM + His I2I CF \& Path} \\
{\ttfamily\small
    Prompt template: You are a movie recommender system now. \\
    Here is the watching history of a user in the past in order:
    \meta{history titles} \\
    To help you generate better recommendation, I will provide you with some collaborative filtering information related to this user's historical watched movies on this platform. For each historical watched movie, I will give you its most frequently co-watched movies by all users. \\
    \meta{His I2I text} \\
    \textbf{I will also provide a reasoning path between several frequently co-watched movie pairs to explain why they are always watched by same user.}\\
    \meta{His I2I path text} \\
    Now there are 20 candidate movies that this user can watch next:
    \meta{candidate titles} \\
    Please rank the 20 candidate movies by measuring the possibilities that this user would like to watch next most, according to the provided watching history and the CF information. Please think step by step.\\
    Your output is only allowed to be a rerank of the candidate list. Do not add movies out of the candidate list.\\
    Please give the results as JSON array (from highest to lowest priority, movie names only):
}
\subsection{Placeholder details}
\noindent\textbf{Placeholder:} {\ttfamily\meta{history titles with feature} and \meta{candidate titles with feature}}\\
{\ttfamily\small
Template: [\meta{Item 1}(genre: \meta{Genre 1}, Publish year: \meta{Year 1}), ..., \meta{Item n}(genre: \meta{Genre n}, Publish year: \meta{Year n})]
}

\vspace{5mm}
\noindent\textbf{Placeholder:} {\ttfamily\meta{His I2I text} and \meta{His Cand I2I text}}\\
{\ttfamily\small
Template: Users who watched \meta{I2I Item A\_1}, their most frequently watched movies in descending order are: \meta{I2I Items B\_1}.\\
...\\
Users who watched \meta{I2I Item A\_n}, their most frequently watched movies in descending order are: \meta{I2I Items B\_n}.
}

\vspace{5mm}
\noindent\textbf{Placeholder:} {\ttfamily\meta{His I2I path text}}\\
{\ttfamily\small
Template: \meta{Entity A\_1} --> (\meta{Relation R\_1}) --> ... --> \meta{Entity B\_1} \\
...\\
\meta{Entity A\_n} --> (\meta{Relation R\_n}) --> ... --> \meta{Entity B\_n}
}

\vspace{5mm}
\noindent\textbf{Placeholder:} {\ttfamily\meta{His U2I item titles} and \meta{His-Cand U2I item titles}}\\
{\ttfamily\small
Template: \meta{U2I Item 1}, ..., \meta{U2I Item n}
}

\end{document}